\UseRawInputEncoding
\documentclass[aip,amsmath,amssymb,reprint]{revtex4-1}

\usepackage{graphicx}   
\usepackage{dcolumn}    
\usepackage{bm}         
\usepackage{color}      
\usepackage{hyperref}   
\usepackage[utf8]{inputenc}

\begin{document}

\preprint{AIP/POF}  


\title{On the interaction of dilatancy and friction in the behavior of fluid--saturated sheared granular materials: a coupled Computational Fluid Dynamics--Discrete Element Method study}



\author{Bimal Chhushyabaga}
\email[]{bchhushyabaga@uh.edu}
\affiliation{Department of Civil and Environmental Engineering, University of Houston}

\author{Behrooz Ferdowsi}
\email[]{behrooz@Central.UH.EDU}
\affiliation{Department of Civil and Environmental Engineering, University of Houston}


\date{\today}

\begin{abstract}
Frictional instabilities in fluid-saturated granular materials underlie natural hazards, including submarine landslides and earthquake initiation. Experimental evidence shows distinct failure behaviors under subaerial and subaqueous conditions due to the coupled influences of mechanical deformation, interparticle friction, and particle--fluid interactions. We use three-dimensional coupled computational fluid dynamics--discrete element method (CFD--DEM) simulations to investigate the collapse and runout of dense and loose granular assemblies in both environments. Parametric analyses demonstrate that pore-pressure evolution controls the failure mode in saturated settings (fast vs slow sliding), consistent with prior laboratory experiments and lattice-Boltzmann--discrete element method simulations: dense assemblies stabilize via dilation, whereas loose assemblies compact rapidly and transiently fluidize. At the mesoscale, we coarse-grain particle-contact statistics and Eulerian fluid fields to define apparent friction and normalized pore pressure, and organize inertial and viscous responses using $\log_{10}(I_n / I_v)$. Spatiotemporal analyses of these coarse-grained fields reveal strain-rate-dependent behavior governed by evolving porosity and effective stress. In both environments, friction in the failure shear zone is rate-strengthening with respect to the inertial number ($I_n$, for dry medium) and viscous number ($I_v$, for fluid-saturated medium). We further utilize the mesoscale stress framework to compare the evolution of pore pressure in the CFD--DEM simulations of subaqueous slope collapse with an analytical solution for the development of the failure front, using inputs derived from numerical triaxial DEM tests on the same assemblies. The analytical model reproduces steady-state excess pore pressures and captures fluid--particle coupling; however, a mismatch near failure onset suggests a role for transient frictional behavior in grain--fluid interactions. These results support the development of physics-based models of natural hazards and advance our mechanistic understanding of saturated granular failure.

\end{abstract}


\maketitle 

\section{Introduction}

Submarine slides and slope failures play a major role in transporting and redistributing sediments from continental shelves to the deep sea, often accompanied by the formation of turbidity currents. As with subaerial landslides, submarine slides occur in multiple modes, each producing distinct flow dynamics and transport potentials. These modes include liquefaction, hydraulic cracking of sediment deposits, and slip along a failure surface \cite{pinet2008surface,blum2010assessing,edwards1995mudflow}. A more complex and comparatively less studied mode is the breaching flow slide, also referred to as a retrogressive slope failure—a slow, gradual frictional slip along a failure surface \cite{you2014mechanics, You2012dynamics, eke2011field, van2002importance}. 

Recent experimental and continuum mechanics–based modeling studies by You et al. \cite{you2014mechanics, You2012dynamics,you2014heterogeneity}
elucidate key processes governing breaching. Using laboratory flume experiments equipped with pore-pressure sensors and sonar imaging systems, they investigated the failure and sliding mechanics of steep subaqueous sediment deposits. Their results indicate that breaching occurs in dense, fine-grained sediments (e.g., sandy silts) due to a transient buildup of negative pore pressure during dilation at the onset of sliding. This negative pore pressure temporarily increases effective stress and shear resistance, arresting the formation of a continuous failure surface; as pore pressure dissipates, slow destabilization resumes, enabling progressive breaching. \citet{you2014mechanics} further demonstrated that breaching spans a spectrum—from full breaching flow slides to dual-mode failures (cyclic alternation between abrupt sliding and slow breaching), and finally to abrupt frictional sliding—as deposit thickness increases for a given slope length. While aspects of these processes have been modeled using continuum approaches for friction–fluid coupled media, the underlying mechanisms—particularly the transient interplay between dilatancy and friction—remain poorly understood under Earth’s near-surface stress conditions. Mathematical and computational frameworks capable of capturing these coupled processes, ranging from discontinuous granular media to continuum scales, are an area of active research and development \citep{Soulaine2024, montella2021two, rauter2021compressible, Guazzelli2018}.

Dilatancy—the tendency of granular materials to change volume under shear deformation\cite{Reynolds1885,Bolton1986,Kabla2009} — plays a central role in granular failure through its direct coupling with pore-pressure evolution \cite{Goren2010}. The interaction between dilatancy and pore pressure governs transitions from stable creep to localized failure and strongly influences landslide mobility and runout \cite{Finnegan2024}. These transitions often occur within shear zones—regions of localized deformation and elevated shear stress that develop in response to external loads or internal structural heterogeneities \cite{Dey2016}. As shear progresses, porosity and grain fabric evolve, promoting frictional weakening \cite{Wan1999}, while pore-pressure feedbacks reduce effective normal stress and amplify instability \cite{Deng2020}, creating conditions conducive to dynamic failure \cite{Iverson2005}. Yet shear zones do not always evolve toward instability; under certain conditions, dilatant hardening can increase slip resistance and stabilize deformation, consistent with observations of slow slip events characterized by quasi-static rupture and field-scale recurrence intervals \cite{Segall2010}. Shear-zone response also depends on initial density: dense zones typically dilate during initial shear deformation, with porosity stabilizing after sufficient shear displacement \cite{Iverson2005,Garagash2003}. In submerged settings, pore fluid  can reduce interparticle friction \cite{Legros2002}, introduce buoyant uplift \cite{Cassar2005}, and decrease the effective weight of the material, collectively enhancing downslope mobility \cite{Vescovi2020}. These factors govern landslide flow dynamics, which may transition among free-fall, inertial, and viscous regimes depending on the Stokes number ($S_t$) and the density ratio ($r$) between the material and the ambient fluid \cite{CourrechduPont2003}.

At the microscale, dense suspensions exhibit a viscoplastic rheology governed by frictional and collisional interactions that evolve with the viscous number, $I_v$ \cite{Boyer2011,Buettner2020}. Shear-zone behavior is further influenced by shear rate \cite{Marone1990}, the prevailing stress state \cite{Iverson2000,Aharonov2002}, and the evolving balance between dilation and compaction. Together, these factors define the stress–strain response and sliding stability of saturated granular systems \cite{Hsiao2017}. Within this framework, the visco-inertial number offers a unified approach for characterizing rheological transitions in mono- and bidisperse granular flows \cite{Cui2020}.

Empirical observations, laboratory experiments, and numerical models collectively show that porosity and dilatancy critically influence frictional strength, pore-pressure evolution, and landslide mobility. Loose soils tend to liquefy abruptly under shear, whereas dense packings dilate, lowering pore pressure and resisting motion \cite{Iverson2000}. Submerged granular-collapse experiments reinforce these trends: loose beds generate rapid, shallow flows, while dense configurations yield slower, more coherent deposits due to enhanced fluid–grain coupling \cite{Rondon2011}. Classical theories, such as Rowe’s stress–dilatancy framework, provide valuable insights into strength evolution under confinement but overlook key parameters such as stress magnitude, void ratio, and initial packing conditions \cite{Wan1999}. Modern constitutive models address these limitations by incorporating state parameters and density-dependent dilatancy to better represent evolving material behavior.

Advances in CFD–DEM modeling have enabled high-resolution investigations of fluid–solid interactions under landslide-relevant conditions \cite{Pham2018}. These simulations demonstrate that pore-pressure evolution controls shear localization: dilation induces suction and stabilizes dense packings, whereas compaction elevates pore pressure and promotes failure in looser configurations. Drainage and permeability strongly modulate these responses, with dilation near drained boundaries increasing effective stress and compaction near impermeable zones reducing shear resistance \cite{Aharonov2013}. CFD–DEM studies also identify distinct flow regimes—grain-inertial, fluid-inertial, and viscous—governed by the Stokes number ($S_t$) and the density ratio ($r$), with important implications for runout, energy dissipation, and deposit morphology \cite{Jing2019}. Semi-resolved models capture vortex structures and fluid-mediated grain interactions, improving predictions of runout extent and deposit architecture \cite{Chen2023}. Bidisperse configurations, especially coarse-over-fine layering, enhance segregation and kinetic energy transfer, promoting longer runouts, whereas reversed or randomly mixed beds tend to inhibit mobility \cite{Huang2025}. The Column aspect ratio also influences vortex formation and segregation intensity; under submerged conditions, increased viscous drag reduces runout distance and delays recompaction \cite{Shademani2021}.

Despite these advances, many of which correspond to the steady-state and final run-out distances of the sediment and granular deposits, the transient coupling among friction, pore pressure, and dilatancy in fluid-saturated granular systems—critical for flow initiation and localization—is not yet sufficiently explored and is not well-studied. This study addresses these gaps using CFD–DEM simulations to investigate the roles of dilatancy, initial porosity, and frictional resistance in the shear deformation of dry and submerged granular systems, with a focus on resolving the evolution of pore-fluid pressure and stress invariants within the failure zone. Two initial packing densities are examined under both subaerial and subaqueous conditions, incorporating Hertzian contact laws for grain–grain interactions. We first present the numerical model setup and boundary conditions, followed by a comparative analysis of failure kinematics and runout distances in the subaerial (dry) and subaqueous (fluid-saturated) cases. We then analyze spatiotemporal variations in coarse-grained stress, strain, and pore-pressure fields to identify emergent failure mechanisms and shear-zone evolution. The simulations are further used to examine how spatial dilation influences pore-pressure evolution and breaching-style failure in both dense and loose granular configurations. We characterize the evolution of friction, porosity, and pore pressure as functions of deviatoric strain rate and dimensionless numbers, including the inertial and viscous numbers. A two-dimensional analytical model for pore-pressure evolution is employed to compare theoretical predictions based on soil mechanics with the CFD–DEM results. To support this comparison, triaxial DEM isotropic unloading tests are conducted to quantify isotropic compressibility and dilation strength, with dilation modeled as a function of mean effective stress and the ratio of major to minor principal effective stresses.

This work advances studies of granular failure processes in fluid-saturated conditions by integrating simulation, analytical comparison, and mesoscale rheological characterization. Specifically, (i) we compare dilatancy–pressure coupling in subaqueous collapse through direct analysis of CFD–DEM–simulated pore-pressure fields and an analytical solution \cite{You2012dynamics}, using DEM-derived elastic and poromechanical parameters (\(E\) and \(\beta\)) \cite{You2012dynamics,you2014mechanics}, thereby linking microstructural deformation to continuum-scale pore-pressure response in the context of breaching and slope-failure hazards \cite{Alhaddad2024-kj}; (ii) we extract local rheological measures and unify inertial and viscous regimes within the combined \(K\)-framework \cite{Trulsson2012}, \(K=I_v+\alpha I_n^2\), enabling consistent comparison of subaerial and subaqueous failure behaviors; and (iii) we perform systematic micro-to-mesoscale upscaling from grain-resolved CFD–DEM data to Representative Volume Elements (RVEs) \cite{Rycroft2009}, allowing coarse-grained definitions of apparent friction and normalized pore pressure that quantitatively relate evolving porosity, effective stress, and sliding dynamics. Collectively, these advances establish a mechanistic basis for understanding how transient fluid–grain interactions govern frictional instability and provide a framework for physics-based modeling of natural hazards involving fluid-saturated granular flows.

\section{Numerical approach: Coupled Computational Fluid Dynamics with the Discrete Element Method (CFD-DEM)}
We use a numerical technique that combines Computational Fluid Dynamics (CFD) with the Discrete Element Method (DEM) to simulate fluid-particle systems. This approach enables detailed modeling of particle-fluid interactions by combining CFD's ability to simulate fluid flows with DEM's capability to track the motion and collisions of individual particles \cite{ElEmam2021}. Below, we provide details of the model setup and computations used in this study. We use the open-source code MFiX to simulate fluid–particle systems, which employs a hybrid Euler–Lagrange framework for multiphase flows, solving the fluid phase on an Eulerian grid and treating particles as discrete Lagrangian entities \cite{MFIXDEM2012,Musser2021}. The CFD and DEM solvers are coupled through momentum exchange between the fluid and particles. Fluid motion is governed by the Navier–Stokes equations and is solved using a predictor–corrector method with an approximate projection scheme. Particle trajectories are determined by Newton’s laws, integrated using a forward Euler scheme with additional collision handling. Interpolation and deposition operators facilitate coupling between the fluid and particle phases.

\subsection{Governing equations}
The governing equations implemented in MFiX describe fluid-phase continuity and momentum conservation, excluding phase changes, chemical reactions, and phenomena such as growth, aggregation, or breakage \cite{Syamlal1993, MFIXDEM2012, Musser2021}. Mass conservation, as expressed in Eq.~\ref{eq:fluid_mass_conservation}, ensures the continuity of the fluid phase by accounting for the volume fraction occupied by particles. Momentum conservation, given in Eq.~\ref{eq:fluid_momentum_conservation}, governs changes in fluid velocity resulting from pressure gradients, viscous stresses, body forces such as gravity, and interactions with particles. Fluid-phase properties are defined at cell centroids, while pressure is defined at cell nodes. The governing equations are solved using a predictor–corrector scheme with an approximate projection to enforce the divergence constraint. Key numerical methods include second-order upwind discretization for convective terms, central differencing for viscous diffusion, and a nodal projection method to maintain incompressibility.

The fluid phase interacts with particles through drag forces and volume displacement, which are handled using interpolation and deposition operations between the Eulerian grid and Lagrangian particles. The fluid-phase volume fraction is denoted by \(\varepsilon_f\), the fluid-phase density by \(\rho_f\), and the volume-averaged fluid velocity by \({v}_f\). Momentum transfer between the fluid and the \(m^{\text{th}}\) solid phase is represented by \({I}_{fm}\), and the total stress in the fluid is captured by the fluid-phase stress tensor \({S}_f\), as defined in Eq.~\ref{eq:fluid_momentum_conservation} and further detailed in Eq.~\ref{eq:fluid_phase_stress_tensor}. This tensor includes contributions from both pressure and viscous stresses, with the viscous stress component explicitly defined in Eq.~\ref{eq:fluid_phase_shear_stress_tensor}. The strain-rate tensor is given by \({D}_f = \frac{1}{2} \left( \nabla {v}_f + (\nabla {v}_f)^T \right)\), where \(\mu_f\) and \(\lambda_f\) are the dynamic and second viscosity coefficients of the fluid phase, respectively.

\begin{equation}
\frac{\partial (\varepsilon_f \rho_f)}{\partial t} + \nabla \cdot (\varepsilon_f \rho_f \mathbf{v}_f) = 0
\label{eq:fluid_mass_conservation}
\end{equation}

\begin{equation}
\frac{D}{Dt}(\varepsilon_f\rho_f\mathbf{v}_f) = \nabla \cdot \mathbf{S}_f + \varepsilon_f\rho_f\mathbf{f} - \sum_{m=1}^{M} \mathbf{I}_{fm}
\label{eq:fluid_momentum_conservation}
\end{equation}
Set~II Interphase coupling:

\begin{equation}
\mathbf{I}_{fm}(\mathbf{x},t)= -\,\frac{1}{V_c}\sum_{p\in V_c}\mathbf{F}^{\,f\rightarrow p}_{\mathrm{hyd}}
\label{eq:Ifm_def_setII_split1}
\end{equation}

\begin{equation}
\mathbf{F}^{\,f\rightarrow p}_{\mathrm{hyd}}
= \mathbf{F}^{\mathrm{drag}}_p \; (+\,\mathbf{F}^{\mathrm{lift}}_p + \mathbf{F}^{\mathrm{vm}}_p\ \text{if enabled})
\label{eq:Ifm_def_setII_split2}
\end{equation}

Here, $V_c$ is the CFD cell volume and $[\mathbf{I}_{fm}]={\rm N\,m^{-3}}$; if a projection kernel $W$ is used, replace $1/V_c$ by $W(\mathbf{x}-\mathbf{x}_p)$ so that $\int_{V_c}\mathbf{I}_{fm}\,dV=-\sum_{p\in V_c}\mathbf{F}^{\,f\rightarrow p}_{\mathrm{hyd}}$. Action–reaction is enforced on the fluid control volume. In this work, we use hindered drag only (\texttt{drag\_type='SYAM\_OBRIEN'} in MFIX-DEM terminology \cite{MFIXDEM2012,MFIX_DEM_Ref}); lift and virtual mass are disabled, while $-\varepsilon_f\nabla p+\nabla\!\cdot(\varepsilon_f\boldsymbol{\tau}_f)$ remain explicit in the fluid and $-V_p\nabla p$ acts on particles \cite{Zhou2010JFM}.

Hydrodynamic lubrication is the near-contact viscous resistance that grows sharply as interparticle gaps shrink \citep{Davis1986}. A scaling analysis \citep{Carrara2019} suggests that explicitly accounting for lubrication forces is not required in our exact regime of saturated granular flows. Following the criteria discussed in Carrara et al. (2019)\cite{Carrara2019} and restated by Breard et al. (2022)\cite{Breard2022}, the effects of lubrication forces become significant if both (i) $\log_{10}(X)>0$ with $X=A v_p/U_T$ and (ii) $\log_{10}(X)/\log_{10}(Y)>1$ with $Y=v_f/(2\alpha U_T)$ and $\log_{10}(Y)>0$. Here, $A$ is a characteristic particle length, $v_p$ is the particle speed relative to the fluid, $U_T$ is the single-grain terminal settling velocity, $v_f$ is a characteristic fluid speed, and $\alpha$ is a dimensionless hindrance factor. We evaluated the parameters over $t\in[0,1.5]$\,s. In the dense case ($\phi=0.579$) $\max\log_{10}(X)=-1.62$ (95th percentile $-2.58$). In the loose case ($\phi=0.444$) $\max\log_{10}(X)=-1.76$ (95th percentile $-2.76$). Thus, condition~(i) never holds, so the joint criterion is never met, irrespective of the log–ratio values. This aligns with prior laboratory experiments \cite{Rondon2011} and two-phase flow modeling studies \cite{Pailha2009} that reproduced reduced subaqueous mobility and pore-pressure feedback without explicit lubrication.

The Saffman lift is a shear-induced transverse force on a small particle with finite slip in a viscous shear flow \citep{Saffman1965}. The classical Saffman lift law was derived for isolated particles \citep{Saffman1965} and is incorporated into the general equation of motion for small rigid spheres in nonuniform flows \citep{Maxey1983}, but the lift coefficient is small, often orders of magnitude lower than drag for isolated spheres \citep{Loth2008}. Similarly, virtual mass effects become significant primarily under strong unsteady slip accelerations or in rapidly accelerating fluids \citep{Auton1988}, and are further illustrated in multiphase flow simulations \citep{Ferrante2004}. These conditions, however, are not representative of our column collapse simulation, where the governing assumptions and parameter ranges are explained in detail in Section~\ref{sec:steps}.  In dense, viscous-dominated granular flows, two-phase simulations that include lift (with added mass and drag) report secondary influence relative to drag and pressure-gradient coupling \citep[e.g.,][]{Chauchat2017SedFoam}. Also, the reduced runout and pore-pressure feedback observed in laboratory experiments \cite{Rondon2011} and two-phase flow modeling studies \cite{Pailha2009} were captured using drag-only coupling, without using Saffman or virtual mass forces. Hence, we implemented hindered drag only, without Saffman lift or virtual mass, with our primary focus on comparing the rheology of saturated granular flows and the role of dilatancy in submerged granular collapses.

\begin{equation}
\mathbf{S}_f = -P_f\mathbf{I} + \boldsymbol{\tau}_f
\label{eq:fluid_phase_stress_tensor}
\end{equation}

\begin{equation}
\boldsymbol{\tau}_f = 2\mu_f\mathbf{D}_f + \lambda_f\nabla \cdot \text{tr}(\mathbf{D}_f)\mathbf{I}
\label{eq:fluid_phase_shear_stress_tensor}
\end{equation}

The Lagrangian particle phase models individual particles and their interactions with both the surrounding fluid and neighboring particles. Each particle is characterized by its position, velocity, and angular velocity, as described by Eqs.~\ref{eq:position_velocity}–\ref{eq:angular_momentum}. Particles are assumed to be spherical, with properties such as volume, mass, and moment of inertia computed from the particle diameter and material density. Their motion is influenced by gravity, fluid--particle interaction forces (e.g., drag), and contact forces arising from collisions with other particles or walls. Particle collisions are modeled using a soft-sphere Hertzian contact model. Time integration is carried out using a forward Euler method, with subcycling applied during collision events. Particle volume is mapped to the Eulerian grid as a continuous volume-fraction field using a transfer kernel to facilitate particle--fluid coupling. This Lagrangian--Eulerian approach enables detailed modeling of particle--particle and particle--fluid interactions, making it suitable for simulating complex multiphase flows. The total drag force acting on the \(i^{\text{th}}\) particle within the \(k^{\text{th}}\) cell and belonging to the \(m^{\text{th}}\) solid phase is denoted by \({F}_d^{(i \in k, m)}\), and the net contact force from other particles is represented by \({F}_c^{(i)}\), both of which contribute to the linear momentum balance as shown in Eq.~\ref{eq:linear_momentum}. The outward-pointing unit normal vector along the particle radius is denoted by \(\eta\), and the total torque acting on the particle is \({T}^{(i)}\). The net force acting on a particle is \({F}_T^{(i)}\), which includes contributions from gravitational acceleration \(g\), drag force \(F_d\), and additional fluid-induced force \(F_f\), as shown in Eq.~\ref{eq:velocity_drag_particle}. The fluid phase is governed by the Navier--Stokes equations as described by Eq.~\ref{eq:N_S}, which accounts for pressure gradients, viscous stresses, body forces such as gravity, and momentum exchange with particles. Within the Eulerian framework, fluid properties, including velocity and pressure, are defined on a fixed grid and evolve according to this equation. Fluid--particle coupling is implemented through source terms in the momentum equations, enabling two-way interactions necessary for capturing realistic multiphase dynamics. Drag forces are computed based on the relative velocity between particles and the fluid.

The particle velocity is \(V_p\), and the fluid velocity is \(u_f\). The fluid pressure is denoted by \(P_f\), and the characteristic relaxation time of a particle is \(\tau_p\). The fluid density \(\rho_f\) and dynamic viscosity \(\mu_f\) are specified in the material properties. The physical properties used in this study are: fluid density \(\rho_f = 1000.0 \, \text{kg/m}^3\), fluid viscosity \(\mu_f = 0.001 \, \text{Pa} \cdot \text{s}\), grain density \(\rho_s = 2500.0 \, \text{kg/m}^3\), and Young’s modulus \(E = 10 \, \text{GPa}\).

\begin{equation}
\frac{dX_p}{dt} = V_p
\label{eq:position_velocity}
\end{equation}

\begin{equation}
m^{(i)} \frac{d{V}^{(i)}(t)}{dt} = {F}^{(i)}_T(t) = m^{(i)}{g} + {F}^{(i\in k,m)}_d(t) + {F}^{(i)}_c(t)
\label{eq:linear_momentum}
\end{equation}

\begin{equation}
I^{(i)} \frac{d{\omega}^{(i)}(t)}{dt} = {T}^{(i)}(t)
\label{eq:angular_momentum}
\end{equation}

\begin{equation}
m_p \frac{dV_p}{dt} = F_d + F_f
\label{eq:velocity_drag_particle}
\end{equation} 

\begin{widetext}
\begin{equation}
\rho_f \left(\frac{du_f}{dt} + u_f \cdot \nabla u_f\right) = -\nabla P_f + \mu_f \nabla^2 u_f + \rho_f g + \sum_{p} \frac{m_p (V_p - u_f)}{\tau_p}
\label{eq:N_S}
\end{equation}
\end{widetext}


\subsection{Steps in numerical approach and parameters} \label{sec:steps}
\subsubsection{Model setup and particle generation}
\begin{figure}
    \includegraphics[width=\columnwidth]{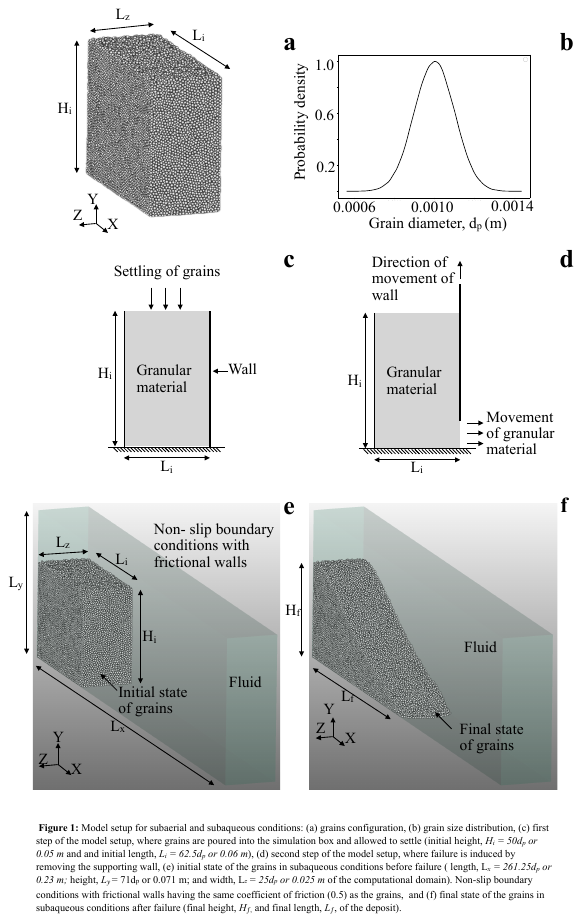}
    \caption{
    Model setup for subaerial and subaqueous conditions: 
    (a) grain configuration, 
    (b) grain size distribution, 
    (c) first step of the model setup, where grains are poured into the simulation box and allowed to settle (initial height, $H_i = 50d_p$ or 0.05 m, and initial length, $L_i = 62.5d_p$ or 0.06 m), with aspect ratio, A  = $H_i$/$L_i$ = 0.8, 
    (d) second step of the model setup, where failure is induced by removing the supporting wall, 
    (e) initial state of the grains in subaqueous conditions before failure (length, $L_x = 261.25d_p$ or 0.23 m; height, $L_y = 71d_p$ or 0.071 m; and width, $L_z = 25d_p$ or 0.025 m of the computational domain). Non-slip boundary conditions with frictional walls having the same coefficient of friction (0.5) as the grains, and 
    (f) final state of the grains in subaqueous conditions after failure (final height, $H_f$, and final length, $L_f$, of the deposit).
    }
    \label{fig:Figure_Model}
\end{figure}

The simulation begins with the generation of particles. As illustrated in Fig.~\ref{fig:Figure_Model}, particles are poured into a simulation box, allowed to settle into equilibrium through cycling, and then trimmed to a specified height to form the final granular column, as shown in Fig.~\ref{fig:Figure_Model}(a). Each column consists of a polydisperse packing of particles, drawn from a normal Gaussian distribution with a mean diameter $d_p$ and a standard deviation $0.1d_p$. The corresponding grain size distribution is presented in Fig.~\ref{fig:Figure_Model}(b). To isolate the effects of initial packing on failure and flow behavior under both dry and fluid-saturated conditions, all columns are prepared with identical geometry: an initial length of $L_i = 62.5d_p$, a height $H_i = 50d_p$, and a width $L_z = 25d_p$. The influence of aspect ratio on flow dynamics has been addressed in previous work \cite{Jing2018}. The computational domain, which houses the column, is a rectangular box with dimensions $L_x = 261.25d_p$, $L_y = 71d_p$, and $L_z = 25d_p$. For fluid-saturated cases, this domain is entirely filled with fluid.

Boundary conditions in the CFD and DEM domains are imposed to replicate a quasi-two-dimensional configuration. In CFD, no-slip conditions are applied on all walls, including the base. For DEM, the boundaries are frictional, with the wall–particle friction coefficient matching that of the grains. Although the DEM framework accommodates a variety of particle shapes and sizes, this study focuses on spherical grains. Relevant particle and fluid properties are provided in Table~\ref{tab:properties}. DEM parameters include a grain density $\rho_p = 2500.0 \, \text{kg/m}^3$, Poisson’s ratio $\nu = 0.23$, and coefficient of restitution $e = 0.2$. Grain--grain and wall--grain Coulomb friction were fixed at $\mu=\mu_w=0.5$, a representative value for quartz sands. In this work, our primary target is to investigate poromechanical coupling; a systematic variation of the grain-grain friction coefficient, $\mu$, and its effect on the failure style and rheology may be examined in future investigations. 

\begin{table}
\caption{Properties of fluid and particles}
\label{tab:properties}
\centering
\begin{tabular}{l c}
\hline
Property & Value \\
\hline \rule{0pt}{2.5ex}
Fluid density ($\rho_f$) & 1000.0~\text{kg/m}$^3$ \\
Coefficient of friction ($\mu$) & 0.5 \\
Fluid viscosity ($\eta$) & 0.001~Pa$\cdot$s \\
Poisson’s ratio ($\nu$) & 0.23 \\
Particle density ($\rho_p$) & 2500.0~\text{kg/m}$^3$ \\
Young’s modulus ($E$) & 10~GPa \\
Coefficient of restitution ($e$) & 0.2 \\
\hline
\end{tabular}
\end{table}

The CFD mesh consists of cells with dimensions $9d_p \times 6d_p \times 3d_p$. Cells were $9d_p \times 6d_p \times 3d_p$ (aspect ratio~3). The minimum grid spacing of $3d_p$ is consistent with unresolved CFD–DEM grid-to-particle sizing guidelines of $\sim2$--$4d_p$ established in verification studies \cite{WangLiu2021,Andrews1996}. A time step of $\Delta t = 10^{-7} \, \text{s}$ ensures tight coupling between the solvers via iterative data exchange. A time step of $\Delta t=10^{-7}\,\mathrm{s}$ is the fluid step; MFIX then subcycles the DEM according to the Hertzian criterion, $\Delta t_{\mathrm{DEM}}=t_{\mathrm{coll}}/N_{\mathrm{coll}}$, where $t_{\mathrm{coll}}$ is computed from the restitution–damping–stiffness relations, Eqs.~31–33, 36–37 using the MFIX–DEM framework for multiphase particulate flows \cite{MFIXDEM2012} and depends on $(E,\nu,\rho_p,d_p,e)$. In our model, $E=10$~GPa, $\nu=0.23$, $\rho_p=2500~\mathrm{kg\,m^{-3}}$, $\bar d_p=1$~mm, and $e=0.2$, giving $\Delta t_{\mathrm{DEM}}\approx10^{-8}\text{–}10^{-7}$~s. Simulations start by initializing both fluid and granular fields. The DEM solver updates grain positions and velocities using fine-scale time integration, resolving collisions and contact mechanics. Simultaneously, interaction forces, such as drag and pressure, are computed from relative motion and volume fractions. These forces are used to update the fluid momentum field, while the revised fluid properties influence subsequent grain dynamics. This loop is executed at each time step, maintaining consistency in the fluid–particle interaction. With the grid and time–stepping choices specified above, the outputs were insensitive to modest mesh coarsening/refinement around the previously stated \(3d_p\) minimum and to halving/doubling both the DEM subcycling factor and the fluid step, consistent with MFIX–DEM verification and validation guidance \citep{MFIXDEM2012,MFIX_DEM_Ref} and time–integration recommendations \citep{MFIX_Time}.

In the second step, particles are packed into the simulation domain. The chosen packing method can significantly influence the resulting packing fraction. In this study, the inlet velocity was varied, and grains were allowed to settle until reaching a stable configuration, as shown in Fig.~\ref{fig:Figure_Model}(c). The packing fraction (\(\phi\)), defined in Eq.~\ref{eq:packing_fraction} as the proportion of the total volume occupied by particles, depends on the number of particles (\(N_{\text{particle}}\)), the volume of an individual particle (\(V_{\text{particle}}\)), and the volume of the unit cell containing the particles (\(V_{\text{unit cell}}\)). It can be adjusted by modifying particle size, shape, or the packing method. For example, introducing smaller particles into the voids of a simple cubic packing can increase \(\phi\). Using this approach, two packing fractions were achieved in this study: 0.579 (dense configuration) and 0.444 (loose configuration). 
To isolate preparation effects from size-distribution effects, we fixed a narrow, truncated-normal particle-size distribution (PSD) ($\sigma_d \approx 0.1\,\bar d_p$). Thus, the dense ($\phi=0.579$) and loose ($\phi=0.444$) states reported here arise primarily from the inlet-velocity compaction protocol rather than differences in polydispersity. We denote $d_p$ as particle diameter, with $\bar d_p$ and $d_{50}$ its mean and median. Similarly, it also limits the increase in computational time that would occur for a few tens of thousands of grains if a log-normal distribution were used \cite{Shire2020}, and avoids introducing small grains for which cohesive forces become significant \cite{CohesiveGranularMaterial2025}. Building on this framework by incorporating wider log-normal distributions would be a valuable direction for future investigations.

\begin{equation}
    \text{Packing fraction},\; \phi = \frac{N_{\text{particle}} V_{\text{particle}}}{V_{\text{unit cell}}}
    \label{eq:packing_fraction}
\end{equation} 

Once the desired packing fraction is achieved, the DEM simulation allows the grains to settle to rest. After settling, the supporting wall is removed to simulate the perturbation or the failure of the landslide, as shown in Fig.~\ref{fig:Figure_Model}(d). The simulation then computes the forces and movements of the particles based on their interactions with each other and with their environment. Figs.~\ref{fig:Figure_Model}(e) and \ref{fig:Figure_Model}(f) illustrate the state and movement of the granular material before and after the failure, respectively.

\subsubsection{Coarse-graining}
We use coarse-graining techniques to transition from the microscale to the mesoscale, employing a least-squares regression to compute the velocity gradient tensor \({L}\) within a Representative Volume Element (RVE) cell, based on the positions and velocities of particles \cite{Rycroft2009}. Granular elements are defined using large-scale, three-dimensional Discrete Element Method (DEM) simulations. The size of the RVE is chosen to match the scale of dynamical correlations in dense granular flows, which emerge over displacements of a few particle diameters as particles transition from superdiffusive to diffusive behavior \cite{Choi2004}, and are typically on the order of 3 to 5 particle diameters \cite{Rycroft2006}. By analyzing the simulation as an ensemble of deforming granular elements, we can locally track material evolution. This framework provides an approximate connection between continuum theories and particle-based simulations, facilitating the study of local relationships between normalized deformation rate, packing fraction, apparent friction, pore pressure, and strain. The input data consist of instantaneous particle positions \({x}\), given by Eq.~\ref{eq:RVEposition}, and velocities \({v}\), given by Eq.~\ref{eq:RVEvelocity}, for all particles within the RVE. The domain is discretized into RVE cells in \((x, y, z)\), and average quantities are computed per bin. For each bin, the deformation gradient tensor \(\mathbf{F}_{ij}\) is calculated as defined in Eq.~\ref{eq:deformation_gradient_tensor}. From \(\mathbf{F}_{ij}\), additional kinematic measures such as the right stretch tensor and strain tensors, Cauchy-Green, Green, and Hencky, are computed.

By minimizing the sum of squared residuals between the observed and predicted velocities as described by Eq.~\ref{eq:RVEvelocity}, we determine both the average velocity and the velocity gradient tensor \(L_{ij}\) in each RVE. The symmetric part of \(L_{ij}\) yields the deformation rate tensor \(D_{ij}\), as defined in Eq.~\ref{eq:RVEstrainrate}. This gives us the components of the strain rate tensor \({D}\). Similarly, the stress tensor \(T_{ij}\) is computed using Eq.~\ref{eq:Stresstensor}, which accounts for all interparticle contact forces within a cell. From this, we extract the normal and shear stress components relevant to the local material response.

The variables used in these equations are defined as follows: \(x_i\) and \(X_j\) are the deformed and reference coordinates, respectively; \(x_{0i}\) is a constant translation vector; \(v_i\) and \(v^0_i\) are the components of the instantaneous and constant velocity vectors respectively; \(L_{ij}\) and \(L_{ji}\) represent the components of the velocity gradient tensor and its transpose, respectively; \(D_{ij}\) is the component of the symmetric deformation rate tensor; \(V\) is the RVE cell volume; \(N\) is the number of particles in the cell; \(N_l\) is the number of contacts for particle \(l\); \(\Delta x_i^{(k,l)}\) is the \(i\)-th component of the vector from the center of particle \(l\) to its \(k\)-th contact; and \(F_j^{(k,l)}\) is the \(j\)-th component of the force on particle \(l\) from its \(k\)-th contact.

\begin{equation}
x_i = \mathbf{F}_{ij} X_j + x_{0i}
\label{eq:RVEposition}
\end{equation}
\begin{equation}
\mathbf{F}_{ij} = \frac{\partial x_i}{\partial X_j}
\label{eq:deformation_gradient_tensor}
\end{equation}

\begin{equation}
v_i = L_{ij} x_j + v^0_i
\label{eq:RVEvelocity}
\end{equation} 
\begin{equation}
D_{ij} = \frac{1}{2} \left( L_{ij} + L_{ji} \right)
\label{eq:RVEstrainrate}
\end{equation}

\begin{equation} T_{ij} = \frac{1}{V} \sum_{l=1}^N \sum_{k=1}^{N_l} \Delta x_i^{(k,l)} F_j^{(k,l)} 
\label{eq:Stresstensor}
\end{equation}

Using the coarse-grained stress tensor, \(T_{ij}\) (Eq.~\ref{eq:Stresstensor}), we compute the shear stress magnitude, \(\tau\) and the effective normal stress, \(\sigma'_n=\sigma_n-p\) (with \(p\) the pore pressure), and obtain the apparent friction, \(\mu_{\mathrm{}}=\tau/\sigma'_n\) and normalized pore pressure, \(p/\sigma'_n\). To explore the nature of failure and flow, dimensionless numbers, specifically the viscous and inertial numbers, are used to distinguish between flow regimes by comparing competing physical forces. The viscous number, \(I_v\), quantifies the relative importance of viscous forces to confining pressure, as defined in Eq.~\ref{eq:viscous_inertial_numbers}. It is commonly applied in systems where fluid resistance dominates particle dynamics, such as dense suspensions or saturated soils. Larger values of \(I_v\) correspond to regimes where viscous effects are more significant. In contrast, the inertial number, \(I_n\), characterizes the ratio of inertial to confining forces in dry or rapid granular flows, where particle collisions and momentum exchange dominate, as given by Eq.~\ref{eq:viscous_inertial_numbers}. Higher values of \(I_n\) indicate stronger inertial effects. In these expressions, \(\eta\) is the dynamic viscosity of the fluid, \(\dot{\gamma}\) is the shear strain rate, \(P\) is the confining pressure, \(d_p\) is the particle diameter, and \(\rho_s\) is the particle density.

\begin{equation}
I_v = \frac{\eta \dot{\gamma}}{P} \quad \text{and} \quad I_n = \frac{\dot{\gamma}\,d_p}{\sqrt{P/\rho_s}}
\label{eq:viscous_inertial_numbers}
\end{equation}

\begin{figure*}[t]
  \centering
  \includegraphics[width=0.64\textwidth] {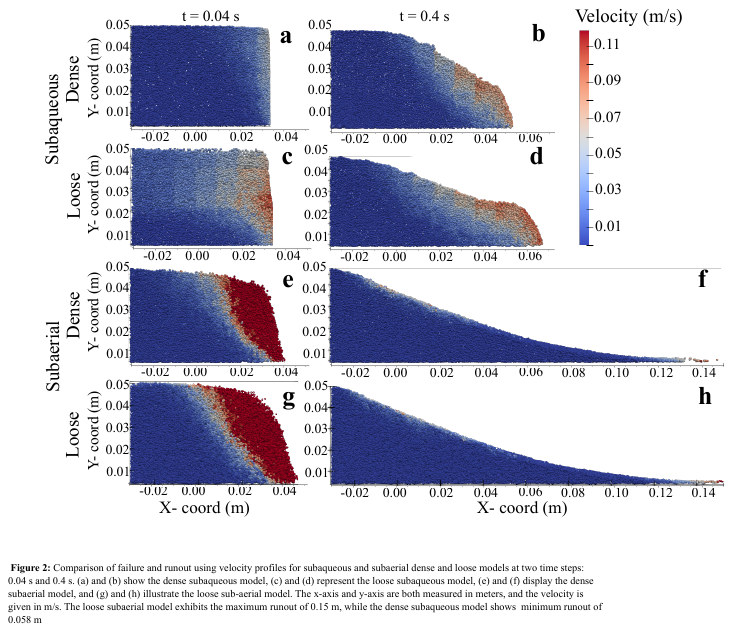}
   \caption{Comparison of failure and runout dynamics using velocity for subaqueous and subaerial models  with dense ($\phi$ = 0.579) and loose ($\phi$ = 0.444) configurations at two time steps: 0.04~s and 0.4~s. Subaqueous model  exhibit slower failure and reduced runout distances compared to subaerial model, due to particle–fluid interactions. Panels (a) and (b) show the dense subaqueous model with a minimal runout of 0.054~m (61\% shorter than the maximum), while panels (c) and (d) represent the loose subaqueous model, which displays greater displacement and a more extended runout of 0.062~m (59\% shorter). Panels (e) and (f) display the dense the subaerial model, characterized by abrupt failure with a runout of 0.15~m (maximum), and panels (g) and (h) illustrate the loose subaerial model, which also exhibits a runout of 0.15~m (maximum).}
    \label{fig:Figure_Comparison_of_failure_and_runout}
\end{figure*}

\section{Mechanism of failure}
The failure mechanisms of granular columns differ significantly between subaerial and subaqueous conditions, with the initial packing fraction playing a crucial role in these behaviors. Subaqueous systems exhibit reduced runout distances and slower failure for both dense and loose configurations due to particle–fluid interactions, with dense configuration being the most stable, a result consistent with previous studies \cite{Zhang2020, Montell2023}. In contrast, subaerial systems fail more rapidly and exhibit greater sensitivity to packing fraction \cite{https://doi.org/10.48550/arxiv.1512.05180}. Loose configurations undergo more pronounced and extended collapses, while dense configurations fail abruptly but with limited spreading. Fig.~\ref{fig:Figure_Comparison_of_failure_and_runout} compares velocity profiles at two time steps (0.04~s and 0.4~s) to illustrate the contrasting failure and runout behaviors across dense and loose configurations in both environments. In subaqueous systems, particle–fluid interactions, such as drag and pore pressure feedback, moderate failure dynamics, leading to slower progression and reduced energy dissipation compared to subaerial cases, although final runout distances may be similar in some cases \cite{Zhang2020}.

The dense subaqueous configuration (Fig.~\ref{fig:Figure_Comparison_of_failure_and_runout}(a) and Fig.~\ref{fig:Figure_Comparison_of_failure_and_runout}(b)) shows minimal runout, with a maximum of 0.054~m, which is approximately 61\% shorter than the maximum runout observed in subaerial conditions. The loose subaqueous configuration (Fig.~\ref{fig:Figure_Comparison_of_failure_and_runout}(c) and Fig.~\ref{fig:Figure_Comparison_of_failure_and_runout}(d)) exhibits a slightly longer runout of 0.062~m, yet still 59\% shorter than its subaerial counterpart. These reductions underscore the dampening influence of fluid on collapse extent. As shearing initiates, dilation occurs due to interlocking grains pushing against one another, increasing the pore pressure and mobilizing peak shear resistance. This interparticle interlocking elevates the internal friction angle and temporarily raises shear strength \cite{Montell2023}. Following this peak, the friction angle decreases, and failure progresses gradually with limited velocity magnitudes throughout the collapse.

\begin{figure}
    \centering
    \includegraphics[width=0.79\columnwidth]{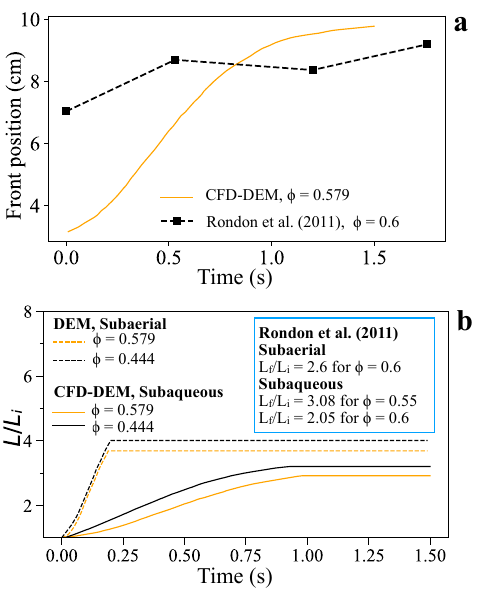}
    \caption{(a) Front-position evolution for a dense subaqueous column collapse. Orange line: CFD–DEM ($\phi=0.579$, $A=0.80$, 3D, laminar). Squares: Rondon \textit{et al.}\ (2011) flume data ($\phi=0.60$, $A\approx0.67$, quasi-2D). Both show rapid advance followed by a plateau; offsets reflect geometry and sidewall effects. (b) Normalized runout, $L/L_i$ for dense \emph{subaerial} and \emph{subaqueous} MFIX–DEM runs compared with Rondon data. Our subaqueous dense result ($L_f/L_i=2.92$ at $\phi=0.579$ and $3.21$ at $\phi=0.444$) falls within the experimental envelope ($L_f/L_i=2.05$–3.08 for $\phi=0.55$–0.60). Subaerial values ($L_f/L_i=3.69$ at $\phi=0.579$ and $4.01$ at $\phi=0.444$) are much larger than quasi-2D flumes ($L_f/L_i=2.6$ at $\phi=0.60$), reflecting reduced sidewall dissipation in 3D setups \cite{Jop2005}. The loose case ($\phi=0.444$) lies below experimental packing ranges and densifies rapidly, preventing pore-pressure-driven mobility. Importantly, the relative ordering (subaerial $>$ subaqueous) remains consistent with experiments \cite{Rondon2011,Polania2024}.}

    \label{fig:Comparision_rondon}
\end{figure}

\begin{figure*}[t]
  \centering
    \includegraphics[width=0.63\textwidth]{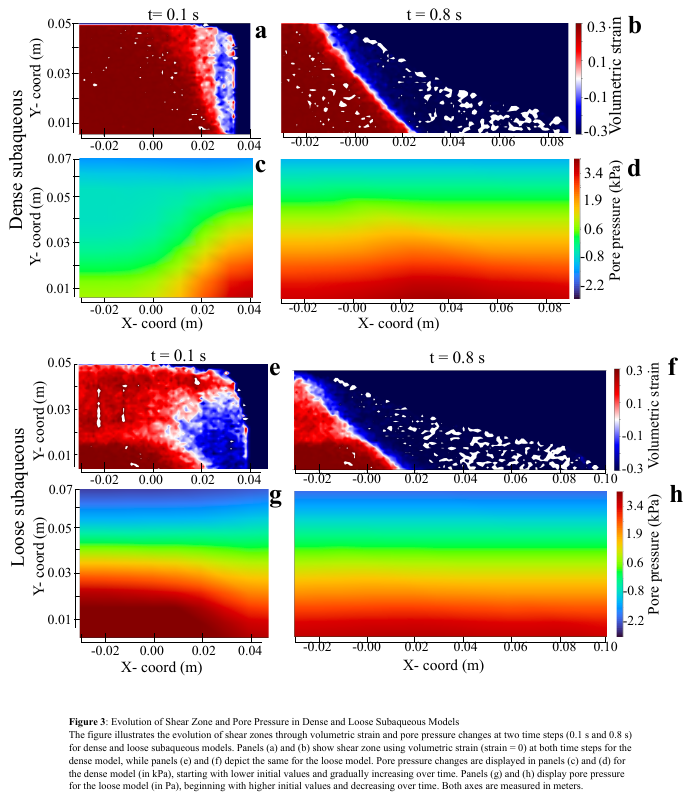}
    \caption{Depiction of shear zone development and pore pressure evolution at two time steps (0.1~s and 0.8~s) for dense ($\phi$=0.579) and loose ($\phi$=0.444) subaqueous models.  Panels (a) and (b) show shear zones, defined by zero volumetric strain within the critical state framework \cite{Wood1991}, for the dense configuration at the two respective time steps, while panels (e) and (f) show the corresponding results for the loose configuration.  Panels (c) and (d) illustrate pore pressure changes in the dense configuration, which begins with lower initial values that increase over time; in contrast, panels (g) and (h) show the loose configuration, which starts with higher initial pore pressures that decrease with continued deformation.}
    \label{fig:Figure_Evolution_of_Shear_Zone_and_Pore_Pressure}
\end{figure*}

\begin{figure}
    \centering
    \includegraphics[width=\columnwidth]{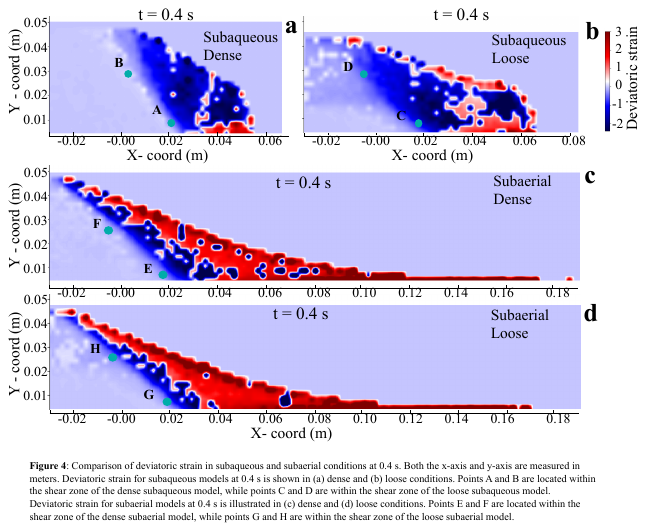}
    \caption{ (a) and (b) show the dense ($\phi$ = 0.579) and loose ($\phi$ = 0.444) subaqueous models, respectively, with points A and B located within the shear zone of the dense configuration, and points C and D within the shear zone of the loose configuration. 
    Panels (c) and (d) illustrate the dense and loose subaerial models, respectively, with points E and F located within the shear zone of the dense configuration, and points G and H within the shear zone of the loose configuration. 
    }
    \label{fig:Figure_Comparison_of_deviatoric_strain}
\end{figure}

\begin{figure*}
    \centering
    \includegraphics[width=0.88\textwidth]{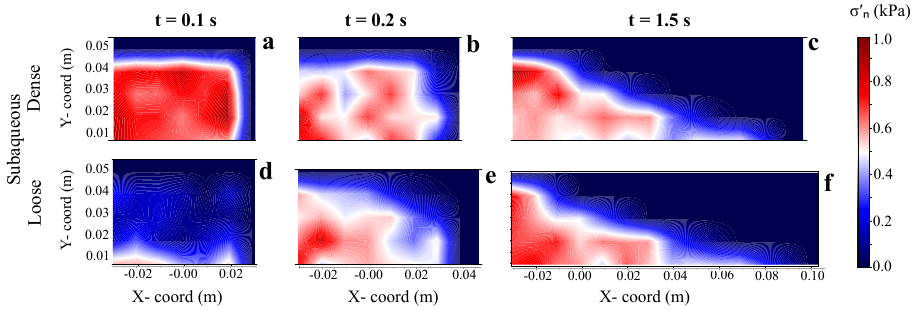}
    \caption{Spatial distribution of normal effective stress $\sigma'_{n}$, for subaqueous configurations at $t=0.1$, $0.2$, and $1.5$ s. Panels (a–c) correspond to the dense packing, and panels (d–f) to the loose packing.}
    \label{fig:Lubrication_potential}
\end{figure*}

\begin{figure*}
    \centering
    \includegraphics[width=0.88\textwidth]{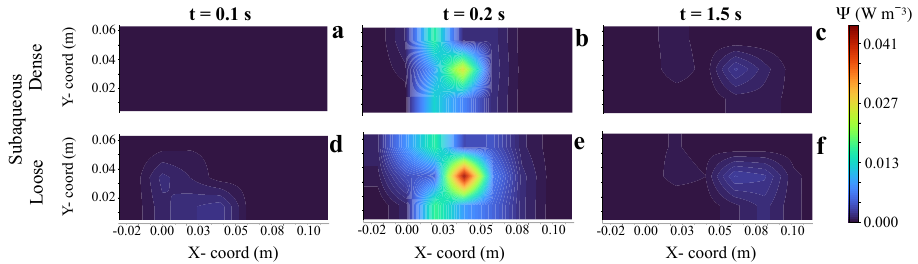}
    \caption{Spatial distribution of viscous dissipation, $\psi$ (W\,m$^{-3}$), for subaqueous configurations at $t=0.1$, $0.2$, and $1.5$ s.
    Panels (a–c) are for dense packing, and panels (d–f) for loose packing.
    Dissipation concentrates near the failure front and is more diffuse in the loose case.
     We compute, $\psi$ using 
    $\psi = 2\mu\,\mathbf{D}:\mathbf{D} - \tfrac{2}{3}\mu(\nabla \cdot \mathbf{u})^{2}$,
    which reduces to $2\mu\,\mathbf{D}:\mathbf{D}$, when $\nabla \cdot \mathbf{u}=0$.}
    \label{fig:Viscous_dissipation}
\end{figure*}

In contrast, the loose subaqueous  configuration experiences greater displacement, as shown in Fig.~\ref{fig:Figure_Comparison_of_failure_and_runout}(c) and Fig.~\ref{fig:Figure_Comparison_of_failure_and_runout}(d). The particles in this configuration are initially loosely packed, resulting in a lower packing fraction and more intergranular void space. This allows for particle rearrangement and contraction during shearing. As shearing progresses, contraction reduces the overall volume and diminishes shear strength due to the loss of effective interlocking, an effect well documented in subaqueous and analogous slope failure studies \cite{Zhang2021,Carey2019}. This internal reorganization leads to more pronounced motion and a longer runout compared to the dense  configuration. Experimental and numerical studies have further shown that loose granular configurations, which are highly sensitive to pore pressure, exhibit increased flow mobility under submerged conditions \cite{Lee2022}. The loose subaqueous  configuration also displays higher velocity, reflecting a more dynamic collapse process. Overall, the contrast between the dense and loose subaqueous configurations highlights the failure behavior governed by the critical role of initial packing density and fluid–particle interactions. While the dense configuration benefits from shear-induced dilation and associated frictional strengthening, the loose configuration, dominated by contraction, increased pore pressure, and reduced interparticle interlocking, undergoes a more mobile and extensive failure.

In subaerial systems, where fluid effects are absent, failure mechanisms are primarily governed by the initial packing fraction of the granular material. The dense configuration, illustrated in Fig.~\ref{fig:Figure_Comparison_of_failure_and_runout}(e) and Fig.~\ref{fig:Figure_Comparison_of_failure_and_runout}(f), demonstrates failure when internal stresses exceed the material's strength. Although the collapse is dilative, meaning that volume increases as particles are sheared, the runout distance remains comparable to that of the loose subaerial case. This limited difference arises from the high initial packing fraction, which enhances interlocking and resistance to deformation \cite{Tapia2013}. In the loose subaerial configuration, shown in Fig.~\ref{fig:Figure_Comparison_of_failure_and_runout}(g) and Fig.~\ref{fig:Figure_Comparison_of_failure_and_runout}(h), exhibits runout, extending to 0.15 meters. The lower initial packing fraction allows particles to rearrange more readily and contract during shearing, leading to increased mobility, which is comparable to the dense configuration. This contraction reduces shear strength and promotes more extensive displacement, as loose grains are less able to resist movement. Experimental findings confirm that such configurations lead to faster and broader collapses, emphasizing the importance of initial packing in governing failure dynamics under dry conditions \cite{bougouin2018granular}. Consequently, the contrast between dense and loose subaerial configurations underscores the critical role of granular structure in determining collapse behavior in the absence of particle–fluid interactions.

\subsection{Comparison of simulated and experimental runouts}
The focus of our study is on investigating the effects of packing fraction and dilatancy on the failure and flow behavior of submerged and saturated granular deposits, and we do not aim to develop a calibrated model of any specific experiment. However, we make an attempt here to compare the behavior of the dense submerged model in our simulations with that observed in experiments on similar materials and dense packing fractions. Fig.~\ref{fig:Comparision_rondon}(a) compares the dense \emph{immersed} front position in our CFD-DEM model against the flume data of Rondon \textit{et~al.} (2011, their Fig.~3A). The CFD–DEM series ($\phi\ = 0.579$, $A = 0.80$, 3D, laminar) and the digitized measurements ($\phi\ = 0.60$, $A = 0.67$) exhibit the same rapid rise followed by a slow approach to a plateau; absolute offsets are attributable to aspect ratio, dimensionality (3D vs.\ quasi-2D), and sidewall effects. Fig.~\ref{fig:Comparision_rondon}(b) reports the normalized runout $L/L_i$ for dense and loose \emph{subaerial} and \emph{subaqueous} cases, alongside panel~(a) front-position comparisons to Rondon \textit{et al.}\ (2011). At $A \approx 0.67$–0.80, Rondon et al.\ measured $L_f/L_i = 2.05$ for subaqueous $\phi=0.60$, $3.08$ for subaqueous $\phi=0.55$, and $2.6$ for subaerial $\phi=0.60$. Our MFIX–DEM runs yield $L_f/L_i = 2.92$ for the subaqueous dense case ($\phi=0.579$) and $3.21$ for the subaqueous loose case ($\phi=0.444$), both consistent with the experimental envelope for $\phi=0.55$–0.60. 

In contrast, the subaerial cases give much larger values, $L_f/L_i = 3.69$ for $\phi=0.579$ and $4.01$ for $\phi=0.444$. These higher values reflect differences in configuration: our fully 3D, flat-bed setup minimizes sidewall dissipation relative to quasi-2D flumes, a factor known to extend runouts in dry conditions \cite{Jop2005}. The loose state in our simulations ($\phi=0.444$) lies below the packing range tested experimentally by Rondon and Polanía and densifies rapidly in the laminar regime, preventing the sustained pore-pressure buildup that can enhance immersed mobility in flumes. Gate-release protocol also matters. In our simulations, the front wall is removed instantaneously at $t{=}0$, yielding an impulsive loss of support and a sharp initial acceleration. In quasi-2D flumes, laboratory collapses are typically initiated by withdrawing a swinging gate or a sliding door with a finite opening time, which transiently supports the column and modifies early shear and pore-pressure transients; this alters the impulse delivered to the front and can shift $L_f/L_i$ at fixed $(\phi,A)$ \cite{Balmforth2005,Rondon2011}. Documented dry-column experiments explicitly note such gate mechanisms (“withdrawing a swinging gate or sliding door”), reinforcing that actuation trajectory and speed are first-order controls on initial momentum and, hence, on runout \cite{Balmforth2005}. Nevertheless, the fundamental ordering (subaerial\,$>$\, subaqueous) is preserved across both simulations and experiments, consistent with the physics of dense column releases.

\subsection{Shear-zone and pore-pressure contrasts in dense vs. loose packings}
There are contrasting behaviors between dense and loose subaqueous configurations in terms of shear zone evolution and pore pressure dynamics. Dense specimens exhibit a slower, progressive build-up of pore pressure under undrained cyclic loading, attributed to their higher resistance to deformation and reduced compressibility at greater densities \cite{Bai2023}. In contrast, loose configurations show broader, more diffuse shear zones accompanied by decreasing pore pressure, as fluid redistributes during contraction and particle rearrangement. This behavior is well captured in multiphase modeling studies, which demonstrate that shear-induced volume changes in loosely packed columns lead to distinct pore-pressure feedback mechanisms, influencing both the timing and extent of collapse \cite{Lee2020}. Experimental studies further underscore the role of fluid: under saturated conditions, reduced flow mobility and increased resistance promote denser frontal accumulation \cite{Tong2022}. Fig.~\ref{fig:Figure_Evolution_of_Shear_Zone_and_Pore_Pressure} provides a detailed visualization of these behaviors at two distinct time steps, 0.1 seconds and 0.8 seconds, highlighting the influence of initial packing density on deformation patterns and pore pressure evolution in fluid-saturated granular columns.

Shear zones in the dense subaqueous configuration  correspond to localized regions of deformation where volumetric strain approaches zero, an interpretation consistent with the critical state framework \cite{Wood1991}. At the early time step of 0.1 seconds, Fig.~\ref{fig:Figure_Evolution_of_Shear_Zone_and_Pore_Pressure}(a), these zones initiate near the base and extend upward in a concentrated pattern. By 0.8 seconds, Fig.~\ref{fig:Figure_Evolution_of_Shear_Zone_and_Pore_Pressure}(b), the shear zones become more pronounced and well-defined, reflecting the progressive nature of shearing during failure. The high initial packing fraction in dense granular configurations enhances frictional resistance, limits deformation, and promotes the development of narrower, more stable shear zones over time \cite{Bai2023,Desrues2004}. These features align with models of dilative deformation in dense, low-permeability granular media, where pore pressure increases due to particle interlocking and restricted fluid redistribution \cite{Petford2003}. Initially, pore pressure remains low in the dense subaqueous configuration because the tightly packed structure inhibits fluid mobility, as shown in Fig.~\ref{fig:Figure_Evolution_of_Shear_Zone_and_Pore_Pressure}(c). As shearing progresses, Fig.~\ref{fig:Figure_Evolution_of_Shear_Zone_and_Pore_Pressure}(d), pore pressure gradually increases, reaching higher values by 0.8 seconds. This rise results from excess pressure generation driven by shearing-induced dilation and the limited ability of fluids to dissipate, characteristics observed in compacted, low-permeability sediments in subduction zones \cite{Nikolinakou2023}. The panels illustrate how pore pressure intensifies in tandem with the development and localization of shear zones.

The shear zones in the loose subaqueous configuration exhibit broader and more diffuse deformation patterns compared to those in the dense configuration. At 0.1 seconds, Fig.~\ref{fig:Figure_Evolution_of_Shear_Zone_and_Pore_Pressure}(e), these zones appear wider and less sharply defined, reflecting the lower initial packing fraction, which promotes particle rearrangement and volumetric contraction. By 0.8 seconds, Fig.~\ref{fig:Figure_Evolution_of_Shear_Zone_and_Pore_Pressure}(f), the deformation zones expand further, illustrating the reduced shear resistance and enhanced mobility in loosely packed configurations. This behavior is consistent with two-phase flow simulations, which show that low-density granular columns develop widespread shear zones early during collapse, accompanied by elevated pore pressures that dissipate as the material contracts \cite{Lee2020}. In such systems, shear bands propagate more readily due to higher initial pore pressures and lower effective stresses. As deformation progresses, these broad zones continue to widen, underscoring the role of initial packing fraction in controlling the kinematic spread of failure. 

In the loose subaqueous configuration, pore pressure initially reaches elevated levels, Fig.~\ref{fig:Figure_Evolution_of_Shear_Zone_and_Pore_Pressure}(g), because the open grain framework allows greater fluid saturation. As shearing advances, Fig.~\ref{fig:Figure_Evolution_of_Shear_Zone_and_Pore_Pressure}(h), pore pressure decreases due to fluid drainage facilitated by contraction and increased permeability behavior that aligns with analytical models describing pressure evolution in contractant granular media \cite{Petford2003}. This reduction in pore pressure supports ongoing deformation and further expansion of the shear zones by illustrating how pore pressure diminishes over time as particles reorganize and the granular structure evolves under submerged conditions.

Deviatoric strain provides critical insights into material deformation under shear, revealing the development of internal failure and strain localization. Comparing dense and loose configurations under both subaqueous and subaerial conditions illustrates how initial packing fraction and fluid presence influence strain deformation and shear resistance. Fig.~\ref{fig:Figure_Comparison_of_deviatoric_strain} presents the spatial distribution of deviatoric strain at $t = 0.4\,\mathrm{s}$, highlighting contrasting shear zone patterns shaped by the interplay between particle arrangement and environmental conditions. In the dense subaqueous configuration (Fig.~\ref{fig:Figure_Comparison_of_deviatoric_strain}(a)), shear localizes into relatively narrow, well-defined zones. The interstitial fluid dampens deformation by reducing interparticle friction, suppressing abrupt strain concentration and promoting a more distributed strain field~\cite{Lee2020,Petford2003}. Points A and B are positioned at the base and middle of the shear zone, respectively, capturing regions of concentrated yet moderated strain due to fluid support.

In contrast, the loose subaqueous configuration (Fig.~\ref{fig:Figure_Comparison_of_deviatoric_strain}(b)) exhibits broader, more diffuse zones of deviatoric strain, indicative of enhanced deformation. The lower packing fraction facilitates particle rearrangement and contraction, resulting in greater mobility and higher strain magnitudes. Shear is less confined, and points C and D, located at the base and center of the shear zone, lie within a more dispersed deformation field. Under dense subaerial conditions (Fig.~\ref{fig:Figure_Comparison_of_deviatoric_strain}(c)), deviatoric strain is sharply localized within narrow shear bands. The absence of pore fluid increases interparticle friction, leading to pronounced strain localization and intense deformation. This observation is consistent with experimental studies on granular collapse under varying fluid conditions, which show that dense packings exhibit distinct flow regimes depending on environmental saturation~\cite{bougouin2018granular}. Points E and F mark the bottom and mid-depth of the shear zone, aligning with regions of high strain typical of dry, densely packed materials.

In the loose subaerial configuration (Fig.~\ref{fig:Figure_Comparison_of_deviatoric_strain}(d)), deviatoric strain remains moderately high but is more broadly distributed than in the dense counterpart. The low initial packing fraction promotes dilation and particle motion, while the lack of fluid amplifies rearrangement without stabilizing resistance. Consequently, points G and H, analogous in location to the other cases, fall within a wider, less confined shear zone, reflecting a more progressive failure pattern and reduced localization intensity. Overall, the figure underscores how fluid presence and initial packing density modulate the spatial extent and intensity of shear zones. Subaqueous models favor distributed deformation, while subaerial conditions promote extensive localization. Loose configurations consistently produce broader deformation zones, emphasizing the central role of initial structure in governing granular shear responses.

We further observe that the dense configurations exhibit pronounced strain localization within narrow shear bands, while loose configurations display broader and less intense shear zones. This contrast stems from lower initial packing fractions and enhanced particle rearrangement in loose configurations. To investigate the evolution of shear zones in more detail, specific points are fixed within the deformation region. These locations enable the analysis of friction, pore pressure, and porosity by tracking associated strain rates and relevant dimensionless numbers, such as the inertial and viscous numbers. Monitoring the temporal evolution of these parameters provides insight into how shear zones behave under different loading conditions. Section~\ref{sec:spatial_temporal} presents results on the spatial and temporal 
distributions of friction, pore pressure, volumetric change, and granular 
rheology at these points in the shear zone.
.

Figures~\ref{fig:Lubrication_potential} and~\ref{fig:Viscous_dissipation} provide additional diagnostics of ambient–fluid effects. Figure~\ref{fig:Lubrication_potential} shows that the effective normal stress, $\sigma'_n\equiv p'$, remains strictly positive across the shear band for both dense and loose packings at all sampled times. The dense configuration largely preserves its initial value for a finite period after failure initiation (gate opening). By contrast, the loose configuration exhibits a pronounced reduction in effective normal stress and a higher initial fluidization potential shortly after gate opening.  Figure~\ref{fig:Viscous_dissipation} shows that viscous dissipation, $\psi$, localizes near the failure front, the area with the highest granular rearrangements, and is more diffuse in the loose case.

\section{Spatial and temporal distribution of friction, pore pressure, and volume change}
\label{sec:spatial_temporal}
\begin{figure*}
    \centering
    \includegraphics[width=0.6\textwidth] {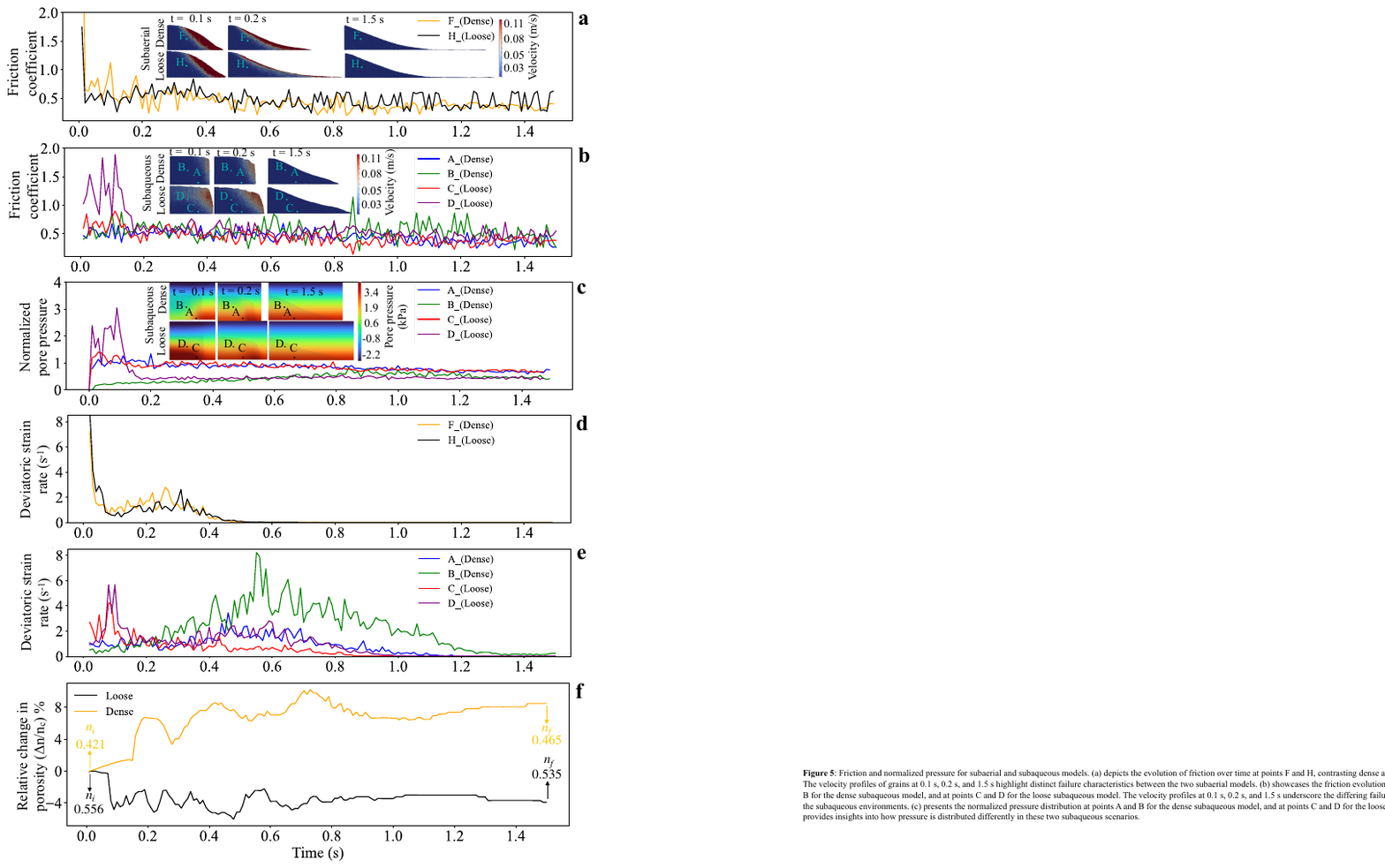}
    \caption{(a) Evolution of mobilized friction coefficient at points F and H, contrasting dense ($\phi$ = 0.579) and loose ($\phi$ = 0.444) subaerial configurations. The mobilized friction is defined as the ratio of shear to effective normal stress, both extracted from the coarse-grained stress tensor \(T_{ij}\) based on interparticle contact forces within each RVE (Eq.~\ref{eq:Stresstensor}). Inset velocity profiles at 0.1 s, 0.2 s, and 1.5 s illustrate distinct failure modes.  (b) Friction evolution at points A and B (dense subaqueous) and C and D (loose subaqueous), using the same stress-based definition.  (c) Normalized pore pressure at the same points, defined as the instantaneous pore pressure divided by effective normal stress in the RVE at each time step, yielding a dimensionless measure of effective stress and fluid–solids interaction.  (d) Deviatoric strain evolution at F and H (subaerial).  (e) Deviatoric strain evolution at A–D (subaqueous), showing contrasting deformation behavior.  (f) Relative porosity change ($\Delta n/n_c$) indicating dilation in the dense ($n_i$ = 0.421, $n_f$ = 0.465) and compaction in the loose ($n_i$ = 0.556, $n_f$ = 0.535), with critical porosity $n_c$ = 0.52.}

    \label{fig:Figure_5_Friction_and_normalized_pressure}
\end{figure*}

Friction at selected points is calculated using coarse-grained stress tensors derived from CFD-DEM simulations. Specifically, the mobilized friction coefficient is defined as the ratio of shear stress to effective normal stress, both extracted from the local stress tensor \(T_{ij}\) (Eq.~\ref{eq:Stresstensor}), which accounts for all interparticle contact forces within an RVE cell. To enable meaningful comparison across different regions and packing conditions, pore pressure values were normalized by the corresponding local normal stress. While normal stress was obtained from coarse-grained stress tensors, pore pressures were sampled at selected locations near the shear zone. This normalization, performed at each time step, yielded a dimensionless pore pressure ratio that facilitates assessment of effective stress evolution and fluid--solid coupling across dense and loose granular systems. The evolution of friction, normalized pore pressure, and deviatoric strain rate over time is shown in Fig.~\ref{fig:Figure_5_Friction_and_normalized_pressure}, providing insight into their dynamic interplay during shear. These time-series plots delineate the transition from transient to steady-state behavior and highlight features such as stick-slip motion and strain localization~\cite{Jessen2024}. We track the time series of the apparent friction, $\mu_{\mathrm{}}(t)=\tau/\sigma'_n(t)$, and the normalized pore-pressure ratio, $p(t)/\sigma'_n(t)$, at labeled shear-band points to quantify transient and relaxation prior to steady state.

\begin{figure}
  \centering
  \includegraphics[width=\columnwidth]{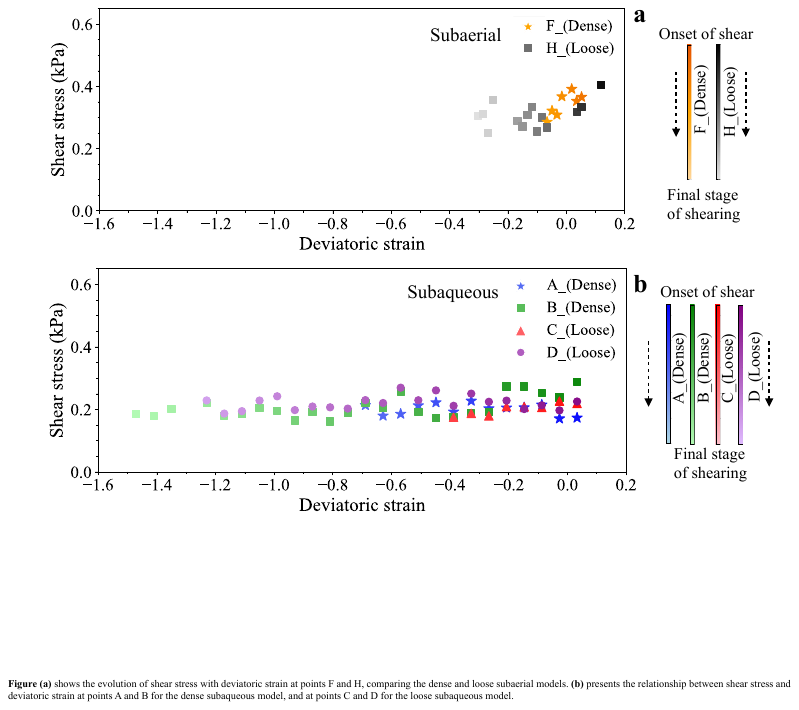}
  \caption{Evolution of shear stress as a function of deviatoric strain for dense and loose granular packings under subaerial and subaqueous conditions. The color gradient from dark to light represents the temporal evolution from the onset of failure to the final stage of deformation. (a) Comparison between the dense (point F) and loose (point H) subaerial models, highlighting differences in mechanical response during forward and reverse shearing. (b) Comparison between the dense (points A and B) and loose (points C and D) subaqueous models, illustrating the influence of pore pressure and packing density on the evolution of shear stress.}
  \label{fig:Figure_shear_deviatoric_strain2}
\end{figure}

Under subaerial conditions, loosely packed granular materials initially exhibit a slightly lower friction coefficient compared to the densely packed model during the onset of failure or early transient stages (Fig.~\ref{fig:Figure_5_Friction_and_normalized_pressure}(a), point H). As shearing progresses, the friction coefficient gradually decreases and stabilizes, reflecting a transition to steady-state behavior characterized by stick-slip motion. This evolution is driven by progressive particle rearrangement and compaction, leading to the development of force chains and increased interparticle contacts~\cite{Jessen2024}. In contrast, densely packed subaerial models exhibit slightly higher friction coefficients at failure, which remain relatively constant as the system enters a steady-state regime that also displays stick-slip characteristics (Fig.~\ref{fig:Figure_5_Friction_and_normalized_pressure}(a), point F). The higher friction in dense packings reflects their more rigid structure, where limited particle mobility reduces the extent of rearrangement and produces a more consistent frictional response~\cite{Huang2024}. Further insight into the mechanical behavior of these configurations is provided by the evolution of shear stress with deviatoric strain, shown in Fig.~\ref{fig:Figure_shear_deviatoric_strain2}(a). For the densely packed model (point F), shear stress varies between approximately 0.3 and 0.4~kPa as deviatoric strain ranges from 0.08 to --0.08, indicating a relatively strong and symmetric response during forward and reverse shearing. In the loosely packed case (point H), shear stress spans from about 0.28 to 0.41~kPa over a broader strain interval of 0.15 to --0.3, reflecting a more compliant response with greater particle rearrangement. These variations highlight the contrasting shear resistance and deformation characteristics between loose and dense subaerial packings.

Deviatoric strain rate (Fig.~\ref{fig:Figure_5_Friction_and_normalized_pressure}(d)) further highlights these differences. Dense configurations show lower strain rates at the start, indicative of constrained particle motion, whereas loose configurations display elevated strain rates during the transient phase due to extensive rearrangement, which gradually declines over time~\cite{Shademani2021}. These strain rate trends correspond with frictional behavior: high strain rates align with low friction in loose packings, while low strain rates are associated with high friction in dense packings (Fig.~\ref{fig:Figure_5_Friction_and_normalized_pressure}(a), Fig.~\ref{fig:Figure_5_Friction_and_normalized_pressure}(d)). These patterns are consistent with experimental findings that associate increased friction with greater particle angularity, broader grain-size distributions, and enhanced surface contact, features characteristic of dense, crushed granular materials~\cite{Tanikawa2025}.

Pore pressure plays a critical role in controlling the mechanical response of granular materials under subaqueous conditions by altering the effective stress and thereby influencing deformation and failure mechanisms~\cite{Beeler2000}. In loosely packed materials, larger void spaces and greater compressibility facilitate a rapid buildup of pore pressure during shearing~\cite{Do2023}. This is characterized by a sharp initial rise in normalized pore pressure, followed by gradual stabilization as particles rearrange and fluid redistributes through the porous structure (Fig.~\ref{fig:Figure_5_Friction_and_normalized_pressure}(c), points C and D). In contrast, densely packed systems initially generate negative pore pressures due to limited pore space and low compressibility, which slows pressure buildup and results in a steadier, more gradual increase (Fig.~\ref{fig:Figure_5_Friction_and_normalized_pressure}(c), points A and B). Despite these differences, both configurations ultimately stabilize as pore pressure equilibrates.

These contrasting pore pressure dynamics strongly influence the frictional response. In loosely packed systems, elevated pore pressures reduce effective normal stress, weakening interparticle contacts and promoting particle rearrangement. This results in a transiently higher friction coefficient in loose packings (Fig.~\ref{fig:Figure_5_Friction_and_normalized_pressure}(b), points C and D) compared to dense packings. Shear stress in these regions ranges from 0.15 to 0.25~kPa over a deviatoric strain interval of approximately 0.02 to --1.22 (Fig.~\ref{fig:Figure_shear_deviatoric_strain2}(b), points C and D). As shearing progresses and fluid drains from the pore space, pore pressure dissipates, causing the friction coefficient to decline and eventually stabilize at a steady-state value comparable to dense subaerial conditions~\cite{Niu2024}.

Conversely, dense subaqueous packings exhibit lower initial friction (Fig.~\ref{fig:Figure_5_Friction_and_normalized_pressure}(b), points A and B) due to suction generated by negative pore pressures (Fig.~\ref{fig:Figure_5_Friction_and_normalized_pressure}(c), points A and B) during dilative deformation . This increases effective stress and reinforces interparticle contacts, limiting particle mobility. Shear stress in these regions ranges from 0.15 to 0.3~kPa as deviatoric strain evolves from 0.02 to --1.45 (Fig.~\ref{fig:Figure_shear_deviatoric_strain2}(b), points A and B)), reflecting a steady buildup of shear resistance and the formation of force chains. These mechanical responses are mirrored in strain localization trends. Loose packings show higher deviatoric strain rates during the early stages of deformation, driven by enhanced particle rearrangement and compaction~\cite{Jessen2024}, whereas dense systems display lower strain rates due to dilative hardening and structural stability (Fig.~\ref{fig:Figure_5_Friction_and_normalized_pressure}(e)). Together, these observations underscore the critical influence of initial packing and pore fluid feedback in shaping both the frictional and kinematic behavior of subaqueous granular materials.

The relative change in porosity ($\Delta n/n_c$) provides critical insight into the volumetric behavior of granular materials during shear deformation, as illustrated in Fig.~\ref{fig:Figure_5_Friction_and_normalized_pressure}(f). It reveals distinct trends of dilation in dense configurations and compaction in loose ones. In the dense configuration, for example, porosity increases from an initial value of $n_i = 0.421$ to a final value of $n_f = 0.465$, indicating dilation. This behavior arises from particle rearrangement and the formation of void spaces, as the material resists further compaction under shear. The critical porosity ($n_c = 0.52$), derived from poromechanics principles, serves as a threshold beyond which dilation dominates the volumetric response~\cite{Shi2021}.

Conversely, the loose configuration exhibits compaction, with porosity decreasing from an initial value of $n_i = 0.556$ to a final value of $n_f = 0.535$, as shown in Fig.~\ref{fig:Figure_5_Friction_and_normalized_pressure}(f). This reduction results from the collapse of void spaces and an increase in particle contacts under shear stress, leading to a denser and more stable configuration. Comparing these behaviors highlights the influence of initial packing density on dilation and compaction mechanisms~\cite{Shi2021}. Dense materials resist further compression and tend to dilate, whereas loose materials undergo compaction as they transition toward a more tightly packed state~\cite{Kruyt2016}. These trends underscore the critical role of porosity evolution in governing shear localization and the overall mechanical response of granular materials~\cite{Kothari2017}.

\begin{figure*}[t]
  \centering
  \includegraphics[width=\textwidth]{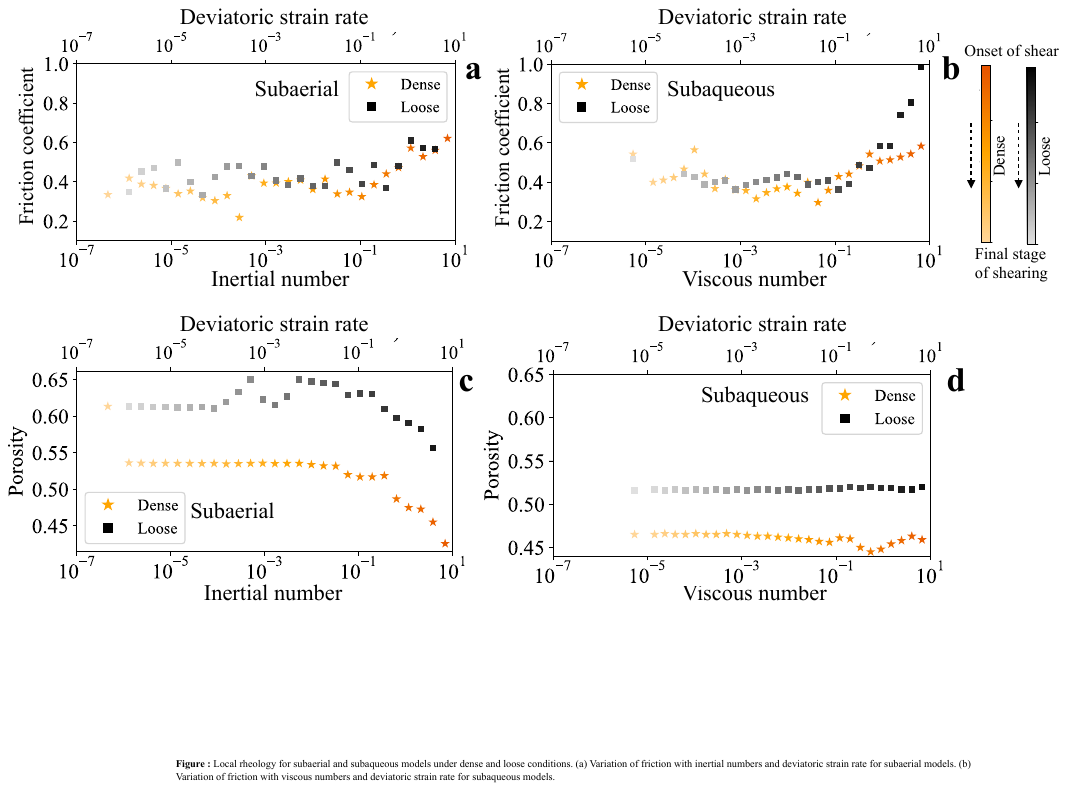}
  \caption{Local rheology for subaerial and subaqueous under dense and loose conditions. The color gradient from dark to light represents the temporal evolution from the onset of failure to the final stage of deformation. (a) Variation of the friction coefficient with inertial number ($I_n$) and deviatoric strain rate for subaerial models, showing rate-strengthening behavior. (b) Variation of the friction coefficient with viscous number ($I_v$) and deviatoric strain rate for subaqueous models, also indicating rate strengthening, with higher frictional values compared to subaerial cases. (c) Variation of porosity with inertial number ($I_n$) and deviatoric strain rate for subaerial models. (d) Variation of porosity with viscous number ($I_v$) and deviatoric strain rate for subaqueous models.
  \label{fig:Figure_Friction_Rate3}}
\end{figure*}

The results above pertain to non-cohesive, quartz-sand–like grains, focusing on the coupling between friction, pore pressure, and tendency for volume-change. Many natural submerged sediments include clayey and silt-sized particles that are known to demonstrate cohesive interactions. The presence of such cohesive interactions is expected to suppress dilation, raise peak friction \citep{Herminghaus2005} at the onset of deformation of materials, shorten runout \citep{Zhu2022}, and subaqueously reduce effective permeability and slow pore-pressure dissipation \citep{Vowinckel2019}. Cohesion also tends to  narrow shear bands in dense flows~\cite{Zhao2023}. Grain-resolved studies report reduced mobility and altered excess-pressure evolution in cohesive beds \cite{Vowinckel2019,Zhu2022}. A cohesive extension of our study is planned as future work, focusing on the systematic exploration and comparison of failure dynamics with cohesive-bed benchmarks \citep{Vowinckel2019}.

\section{Granular rheology in shear zone}
The granular rheology of subaerial and subaqueous models under dense and loose conditions provides critical insights into the interplay between friction, strain rate, and dimensionless parameters such as the inertial number (\(I_n\)) and viscous number (\(I_v\)). These relationships reveal how deformation characteristics and mechanical behavior evolve under varying shear conditions. Fig.~\ref{fig:Figure_Friction_Rate3} illustrates these trends, highlighting rate-strengthening behavior in both environments, driven by distinct mechanisms and frictional responses.

\begin{figure}
  \centering
  \includegraphics[width=\columnwidth]{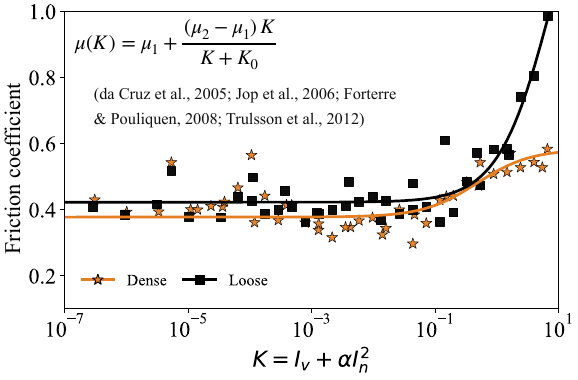}
  \caption{Unified rheology across subaerial (inertial number \(I_n\)) and subaqueous (viscous number \(I_v\)) conditions using the combined number \(K=I_v+\alpha I_n^2\). Friction coefficient \(\mu\) as a function of \(K\); the curve follows Eq.~\ref{eq:muK}. Fitted parameters are reported in the text.}
  \label{fig:Figure_mu_phi_K2}
\end{figure}

\begin{figure}
  \centering
  \includegraphics[width=\columnwidth]{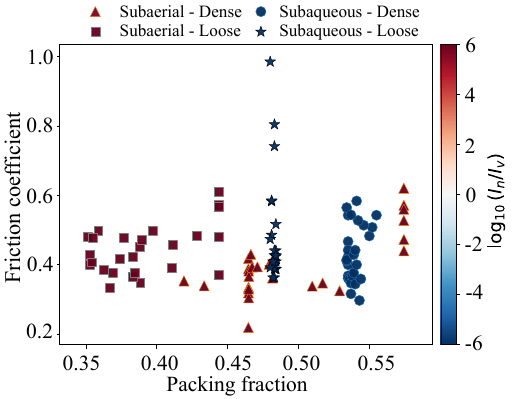}
  \caption{Friction–packing trends across subaerial and subaqueous flows. Symbols show friction \(\mu\) versus packing fraction \(\phi_s\), colored by \(x=\log_{10}(I_n/I_v)\), where \(I_n\) and \(I_v\) are the inertial and viscous numbers, respectively.}
  \label{fig:Figure_N_V_mu_phi}
\end{figure}

\begin{figure}
  \centering
  \includegraphics[width=\columnwidth]{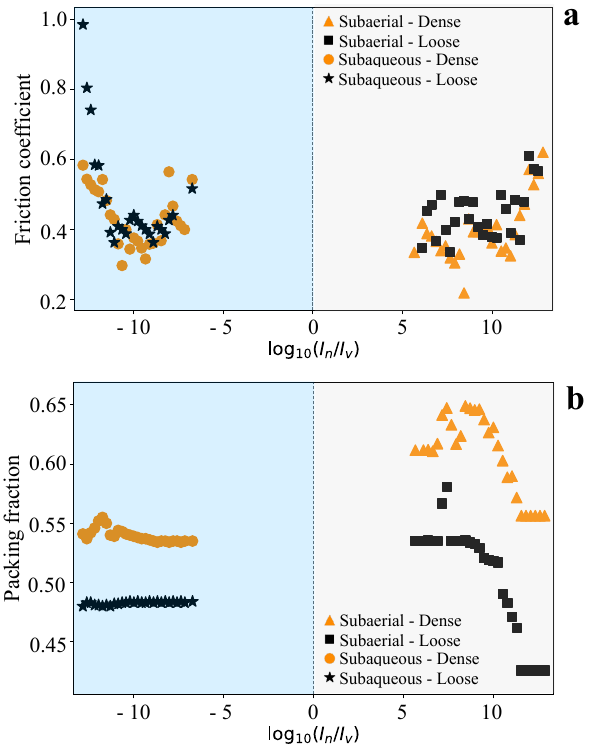}
  \caption{(a) Friction coefficient \(\mu\) versus \(x=\log_{10}(I_n/I_v)\) for subaerial and subaqueous flows. (b) Packing fraction \(\phi_s\) versus \(x\). The vertical dashed line marks the crossover \(I_n=I_v\) (\(x=0\)). Shading indicates regimes: inertial (\(x>0\), light gray) and viscous (\(x<0\), light blue).}
  \label{fig:Figure_N_V_mu_phi2}
\end{figure}

\subsection{Dry condition}

Frictional behavior in subaerial granular systems is strongly influenced by strain rate and the inertial number (\(I_n\)), reflecting the dominant role of dry interparticle interactions. Dense subaerial materials exhibit a modest but consistent increase in friction with rising strain rate (Fig.~\ref{fig:Figure_Friction_Rate3}(a)), characteristic of a rate-strengthening response, with the color gradient representing the temporal evolution from the onset of failure (dark) to the final stage (light) of deformation. This behavior stems from tightly packed grain assemblies that promote the formation of persistent force chains and robust interparticle contacts, thereby enhancing resistance to shear deformation. 

In contrast, loosely packed subaerial materials exhibit lower friction at low strain rates because grain contacts are sparse and short-lived. As the strain rate increases, friction rises more sharply, driven by progressive particle rearrangement, localized compaction, and the development of new contact networks throughout deformation (Fig.~\ref{fig:Figure_Friction_Rate3}(a)). The color gradient indicates time, from the onset of failure (dark) to the final stage (light). This rate-strengthening evolution reflects the reorganization of the granular framework under increasing shear.

Plotting friction against the inertial number \(I_n\) shows that both dense and loose granular packings increase monotonically (Fig.~\ref{fig:Figure_Friction_Rate3}(a)). The response of loose packings is more gradual, with a pronounced rise only at higher \(I_n\). These patterns highlight the controlling role of packing density in strain localization, contact mechanics, and the frictional evolution of dry granular systems.

While porosity in granular media is generally expected to vary with strain rate due to the competing effects of compaction and dilation, it can remain relatively stable over certain regimes. This constancy arises when the system achieves a balance between structural compression and dilation, or when strain rates are too low to significantly disrupt the grain arrangement. In the dense configuration, porosity remains nearly constant for inertial numbers below \(10^{-1}\) (Fig.~\ref{fig:Figure_Friction_Rate3}(c)). \citet{Parez2021} reported a similar trend, observing porosity stability across \(I_n\) values ranging from \(10^{-6}\) to \(10^{-2}\), consistent with our findings. Likewise, \citet{Amarsid2024} found that porosity remains largely unaffected at low inertial numbers, further corroborating these observations for both dense and loose conditions.

\subsection{Fluid-saturated condition}

Subaqueous granular rheology is set by the coupled interactions between solid grains and the surrounding fluid. In these systems, friction reflects both interparticle contact forces and hydrodynamic interactions, yielding greater complexity than in dry granular media. Loose subaqueous materials exhibit a strain-rate-dependent response with two regimes (Fig.~\ref{fig:Figure_Friction_Rate3}(b)). At the onset of shearing, increasing strain rate drives pore-pressure buildup, which elevates friction via fluid drag and the rapid formation of reinforced contact networks. At lower strain rates—typically in the later stages of shearing—fluid lubrication dominates, facilitating particle rearrangement and reducing friction. The resulting rate-strengthening response underscores the central role of fluid-mediated forces in controlling deformation in loosely packed systems.

Dense subaqueous packings respond differently: the friction coefficient decreases with increasing strain rate at shear onset (Fig.~\ref{fig:Figure_Friction_Rate3}(b)), driven by dilation and the attendant negative pore pressure (suction). With further increases in rate, fluid drag becomes increasingly influential and pore-pressure effects intensify, reflecting the evolving coupling between the solid framework and the pore fluid. When friction is plotted against the viscous number \(I_v\), dense packings approach a plateau at high \(I_v\) with reduced variability, suggesting that viscous forces moderate the frictional response under steady fluid--particle interactions. In contrast, loose packings show a more pronounced rise in friction with increasing \(I_v\), owing to intensified fluid drag and pore-pressure buildup.

Although porosity in fluid-saturated granular materials is often expected to vary with strain rate, we observe that it remains nearly constant over a moderate range of rates (Fig.~\ref{fig:Figure_Friction_Rate3}(d)). This invariance arises when compaction and dilation reach a dynamic balance, or when the imposed rates are too low to meaningfully perturb the microstructure. Consistent with these trends, \citet{Amarsid2024} reported that porosity—parameterized by the viscous number \(I_v\)—is largely insensitive at low \(I_v\).

A direct comparison of dry and saturated systems highlights the governing roles of fluid presence and initial fabric in the shear response: subaerial deformation is dominated by interparticle contact mechanics, whereas subaqueous deformation reflects coupled interactions among fluid drag, pore-pressure dynamics, and microstructural reorganization.

Building on these trends, we quantify the rheology by fitting data spanning the inertial and viscous regimes within the combined $K$-framework~\citep{Trulsson2012} (Fig.~\ref{fig:Figure_mu_phi_K2}). The friction law is
\begin{equation}
\mu(K)=\mu_1+\frac{(\mu_2-\mu_1)\,K}{K+K_0},
\quad \text{with } K=I_v+\alpha I_n^2,
\label{eq:muK}
\end{equation}
where \(\mu_1\), \(\mu_2\), \(K_0\), and \(\alpha\) are fitting parameters.

Friction (Fig.~\ref{fig:Figure_mu_phi_K2}) is well described by the rate-strengthening law in Eq.~\eqref{eq:muK} across both subaerial and subaqueous regimes. To interpret this fit, we examine the relation between friction and packing fraction (Fig.~\ref{fig:Figure_N_V_mu_phi}) as a function of the inertial-to-viscous ratio, \(x=\log_{10}(I_n/I_v)\). The data delineate inertial- and viscous-dominated regimes in both environments, with a clear transition near \(x=0\) (\(I_n/I_v=1\)). Plotting friction (Fig.~\ref{fig:Figure_N_V_mu_phi2}(a)) and packing fraction (Fig.~\ref{fig:Figure_N_V_mu_phi2}(b)) against \(x\) cleanly separates the two regimes and reveals how the rheology evolves across subaerial and subaqueous conditions.

To directly compare the inertial and viscous contributions, Figs.~\ref{fig:Figure_N_V_mu_phi2}(a,b) plot the friction and packing fraction against \(x=\log_{10}(I_n/I_v)\). For \(x>0\) (inertial regime), \(I_n\) dominates; for \(x<0\) (viscous regime), \(I_v\) dominates. In the inertial regime, friction increases while packing fraction decreases with shear rate, consistent with classical dry-granular rheology. In the viscous regime, packing fraction varies only weakly with rate, reflecting fluid-drag control and a diminished role for dilatancy. Together with Fig.~\ref{fig:Figure_N_V_mu_phi} (friction versus packing fraction colored by \(x\)), these plots show that the material state spans a continuum between inertial- and viscous-dominated rheologies. This viewpoint naturally connects to the unified \(K\)-framework~\citep{Trulsson2012} (Fig.~\ref{fig:Figure_mu_phi_K2}) and rationalizes the convergence of dense and loose flows when scaled by the dominant stress contribution. Overall, the frictional state is captured by the balance between inertial and viscous stresses, supporting the applicability of the unified rheological framework.

To further interrogate these trends, we compare our simulations and fitted parameters with those of \citet{Trulsson2012}, who calibrated a similar rheology using a two-dimensional DEM of disks coupled to a continuum fluid in a simple–shear cell with periodic \(x\)–boundaries, run at controlled normal stress \(P_p\) and imposed mean shear velocity. By contrast, our analysis employs three-dimensional CFD–DEM column collapses of fluid-saturated beds with dense and loose initial packings, in which shear bands emerge naturally and stresses evolve transiently from grain motion rather than being servo–controlled. These differences in dimensionality, loading protocol, and stress control plausibly account for the deviations in our fitted \(\mu(K)\) relative to simple–shear geometries.

Similarly, \citet{Trulsson2012} report simple-shear simulations in which the inertial-to-viscous number ratio spans \(I_n/I_v \in [10^{-1},10^{4}]\) (i.e., \(\log_{10}(I_n/I_v)\in[-1,4]\)). By contrast, our data span \(I_n/I_v \in [10^{-6},10^{6}]\) (i.e., \(\log_{10}(I_n/I_v)\in[-6,6]\)), extending the accessible range by seven decades (Fig.~\ref{fig:Figure_N_V_mu_phi}) and thereby capturing a broader spectrum of transient shear-zone dynamics during failure.

In the framework of \citet{Trulsson2012}, viscous and inertial contributions are unified with a fitted coefficient \(\alpha=0.635\pm0.009\) and the constitutive form of Eq.~\eqref{eq:muK}, yielding \(\mu_1=0.277\pm0.001\), \(\mu_2=0.85\pm0.01\), and \(\sqrt{K_0}=0.29\pm0.01\). By contrast, fitting the same unified friction law (Eq.~\ref{eq:muK}) to our data gives \(\mu_1=0.378\pm0.018\), \(\mu_2=0.583\pm0.055\), \(\sqrt{K_0}=0.69\pm0.59\), and \(\alpha=1.232\pm2.086\), with coefficients of determination \(R^2=0.597\) (dense) and \(R^2=0.814\) (loose). Both dense and loose cases are rate-strengthening, with loose systems attaining higher friction coefficients.

\section{Analytical solution for excess pore pressure using poromechanics}
During breaching, a dynamic equilibrium emerges between pore pressure dissipation and continuous dilation \cite{You2012dynamics,Alhaddad2024-kj}. As pore pressure dissipates, the effective stress within the slope deposit increases, stabilizing the sediment. However, for retrogressive failure to persist, this dissipation must be sustained by ongoing dilation. The resulting feedback mechanism, where slope failure induces dilative strain that regenerates negative excess pore pressure, helps maintain the breaching process. Importantly, this dynamic decouples the steady-state pore pressure distribution from the deposit’s intrinsic permeability, revealing an underexplored aspect of sediment failure behavior \cite{You2012dynamics,Alhaddad2024-kj}.

To formalize this equilibrium, the analytical solution developed by \citet{You2012dynamics} describes the pore pressure evolution coupled with dilation during the breaching process. This formulation predicts the steady-state negative pore pressure field, showing that the system can self-regulate pore pressure independently of permeability once dilation-driven feedback is active.

Comparisons with large-scale laboratory experiments and numerical models confirm this feedback mechanism’s role in sustaining breaching and highlight deviations during the transient initiation phase \cite{Alhaddad2024-kj,Alhaddad2023}. The same analytical formulation is applied in this study to evaluate whether the CFD--DEM simulation results for excess pore pressure in both dense and loose subaqueous granular models follow the theoretical predictions, particularly at steady state. By extracting simulation data at various locations and times, we directly compare the evolution and final distribution of pore pressure against the analytical model.

Furthermore, recent hydromechanical models of subduction zone sediments \cite{Nikolinakou2023} demonstrate that shear-induced overpressures, in addition to changes in mean stress, significantly control pore pressure evolution. Although applied in a tectonic context, these findings conceptually reinforce that dilative strain and shear play a dominant role in transient and steady-state pore pressure fields, complementing observations from breaching processes.

\begin{equation}
C_v \frac{\partial^2 \hat{u}}{\partial \hat{x}^2} + v \frac{\partial \hat{u}}{\partial \hat{x}} - \beta v \frac{\partial \hat{\sigma}_3}{\partial \hat{x}} = 0
\label{eq:steady_erosion_rate}
\end{equation}

\begin{equation}
\hat{u}^* = \frac{\delta \beta s_0}{1-\delta} \left( e^{-\eta\hat{x}} - e^{-\frac{v}{C_v} \hat{x}} \right)
\label{eq:excess_pore_pressure}
\end{equation}

Eq.~\ref{eq:steady_erosion_rate} describes pore pressure dissipation through Darcy flow with a moving boundary (first two terms) and continuous dilation effects (third term). The coefficient of consolidation, \(C_v\), combines the effects of permeability, viscosity, and compressibility. \(\hat{u}\) represents the excess pore pressure, \(\hat{\sigma}_3\) is the least principal stress (assumed horizontal), and \(C_v\) is the coefficient of consolidation. The boundary conditions applied in this analytical formulation are critical to reproducing the steady-state pore pressure field. In the Lagrangian reference frame, where the coordinate system moves with the retreating failure front, hydrostatic pore pressure is assumed at both the upstream (breaching front) and downstream (far-field) boundaries \cite{you2014mechanics}. Additionally, no-flow conditions are imposed at the base and at the right boundary of the computational domain, assuming symmetry and minimal lateral drainage. These boundary constraints ensure that pore pressure evolves solely due to dilation-induced stress changes and fluid migration within the domain. The steady-state solution derived under these conditions shows that the excess pore pressure depends only on the dilation potential, effective stress, and retreat velocity, but not on permeability once equilibrium is reached. This insight reveals that boundary-controlled feedback mechanisms can dominate the evolution of excess pore pressure during breaching, consistent with both pore pressure measurements and model predictions \cite{you2014mechanics}. Its solution, given by Eq.~\ref{eq:excess_pore_pressure}, represents \(\hat{u}^*\), the excess pore pressure in the Lagrangian reference frame. The parameter \(\beta\) denotes the relative dilation strength, defined as the ratio between differential stress-induced strain and mean effective stress-induced strain. \(s_0\) represents the far-field horizontal stress. The coefficient of consolidation is defined as \(C_v = \frac{k}{\mu E}\), where \(k\) is the permeability, \(\mu\) is the dynamic viscosity of the pore fluid, and \(E\) is a modulus related to volumetric strain. The coordinate \(\hat{x}\) is defined in the Lagrangian reference frame as \(\hat{x} = x - vt\), where \(v\) is the velocity of the retreating breaching front (Eq.~\ref{eq:The_steady_state_erosion_rate}). \(\delta\) characterizes the ratio between \(v\) and the product of the effective stress decay rate \(\eta\) and the coefficient of consolidation \(C_v\). The parameter \(\eta\) is a constant with units of inverse length, affecting the spatial distribution of excess pore pressure.

\begin{equation}
C_v = \frac{k}{\mu E}
\label{eq:Coefficient_of_consolidation}
\end{equation}

\begin{equation}
k = \frac{\mu \, v_{\text{avg}} \, L}{\Delta P}
\label{eq:permeability}
\end{equation}

\begin{equation}
\beta = 1 + \frac{1}{2E} \frac{d \varepsilon_v}{dq}
\label{eq:Dimensionless_dilation_strength}
\end{equation}

\begin{equation}
E = \frac{d \varepsilon_v}{dp'}
\label{eq:isotropic_unloading_compressibility}
\end{equation}

\begin{equation}
\frac{\partial \hat{\sigma_3}}{\partial \hat{x}} = \eta s_0 e^{-\eta\hat{x}}
\label{eq:Least_principal_stress_change}
\end{equation}

\begin{equation}
v = \delta \eta C_v
\label{eq:The_steady_state_erosion_rate}
\end{equation}

\begin{figure}
    \centering
    \includegraphics[width=\columnwidth]{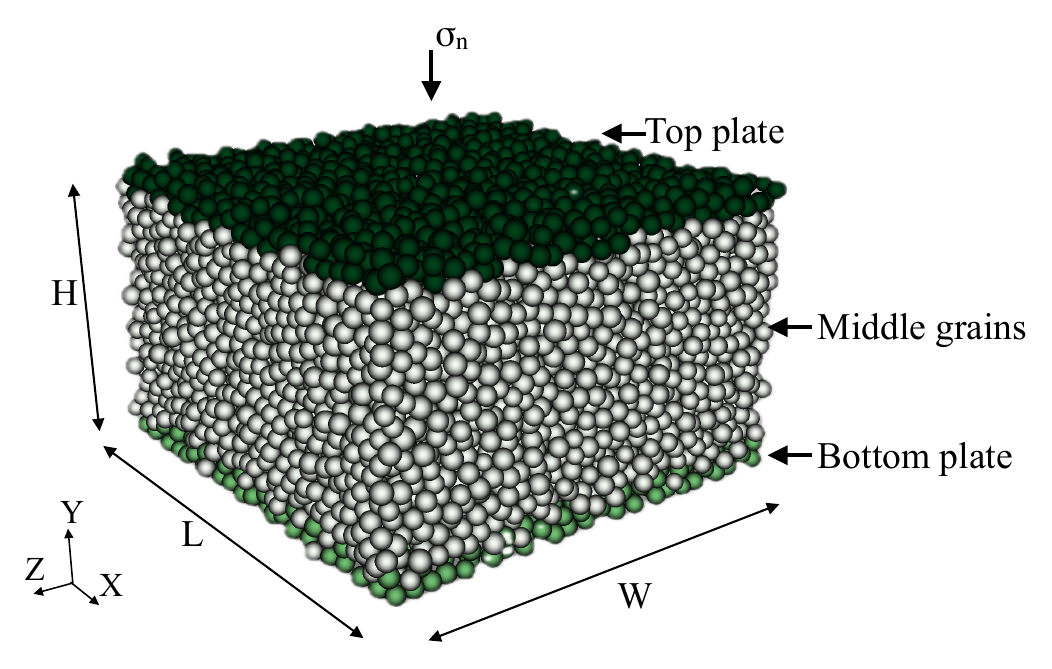}
    \caption{Isotropic unloading compressibility test with applied normal stress \(\sigma_n = 100 \, \text{kPa}\) on the top plate. The specimen dimensions are: \(d_p = 0.001 \, \text{m}\), length (\(L\)) = \(50d_p\), width (\(W\)) = \(50d_p\), height (\(H\)) = \(40d_p\), and plate thickness is \(3d_p\).}
    \label{fig:Figure_Isotropic_unloading_compressibility}
\end{figure}

\begin{figure}
    \centering
    \includegraphics[width=\columnwidth] {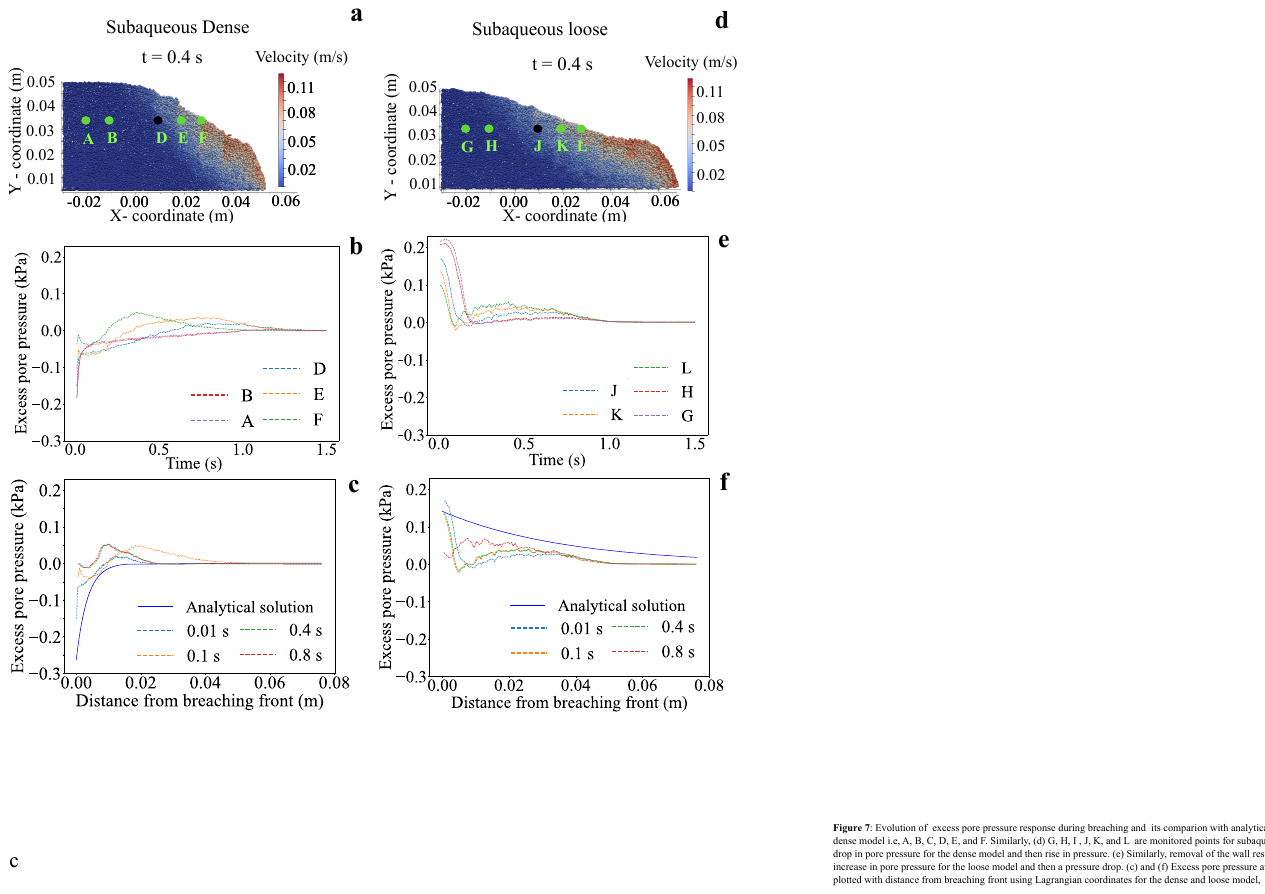}
    \caption{Evolution of excess pore pressure response during breaching and comparison with the analytical solution: (a) Monitored points A, B,  D, E, and F for the subaqueous dense configuration; (d) monitored points G, H, J, K, and L for the subaqueous loose configuration. (b) In the dense configuration, wall removal results in an abrupt drop in pore pressure, followed by a gradual rise as the system stabilizes. (e) In contrast, the loose configuration exhibits an abrupt increase in pore pressure after wall removal, followed by a gradual decline. (c) and (f) Excess pore pressure at time intervals of 0.01 s, 0.1 s, 0.4 s, and 0.8 s (with dashed lines indicating the onset of breaching) is plotted against the distance from the breaching front using Lagrangian coordinates for the dense (point D) and loose (point J) configurations, respectively. The solid line represents the analytical solution. The comparison shows good agreement at steady state for both models: in the dense configuration, pore pressure begins at a negative value, increases, and stabilizes; in the loose configuration, pore pressure starts high, decreases, and reaches a stable value. Transient phases, however, show less agreement with the analytical model, highlighting the complexity of early-stage fluid–particle interactions.
}
    \label{fig:Figure_excess_pore_pressure2}
\end{figure}

For the coefficient of consolidation (\(C_v\)) given in Eq.~\ref{eq:Coefficient_of_consolidation}, \(k\) represents permeability, \(\mu\) is viscosity, and \(E\) is the isotropic unloading compressibility (Eq.~\ref{eq:isotropic_unloading_compressibility}). The compressibility \(E\) is determined as the slope of the volumetric strain \(\varepsilon_v\) versus the mean effective stress \(p' = u_c - u_p\), where \(u_c\) is the cell pressure and \(u_p\) is the pore pressure. The differential stress is denoted by \(q\).

Eq.~\ref{eq:permeability} defines permeability \(k\) as a function of fluid properties and flow path geometry, where \(v_{\text{avg}}\) is the average velocity, \(L\) is the flow path length, and \(\Delta P\) is the pressure difference. The calculated permeability values are \(k = 1.75 \times 10^{-7} \, \mathrm{m}^2\) for the loose configuration and \(k = 0.25 \times 10^{-7} \, \mathrm{m}^2\) for the dense configuration. The evolution of permeability over time at selected points, derived using Darcy’s law, is presented in Supplementary Fig.~S1(a). In Eq.~\ref{eq:Dimensionless_dilation_strength}, \(\beta\) is the dimensionless dilation strength, where \(\beta > \frac{1}{2}\) indicates dilative material. Eq.~\ref{eq:Least_principal_stress_change} describes the least principal stress, which decays exponentially with distance from the breaching front, explaining localized pore pressure changes during continuous dilation. During breaching, the volumetric strain rate equals the water flux into the deposit, and strain is proportional to the minimum pore pressure. The parameter \(s_0\) represents the least principal stress as \(x \to \infty\), and \(\eta\) defines the rate of stress decay.

This interaction establishes a steady-state condition where pore pressure dissipation balances continuous dilation, resulting in a steady pore pressure that becomes independent of permeability. To measure isotropic unloading compressibility and relative dilation strength, we conduct a triaxial DEM isotropic unloading compressibility test (Fig.~\ref{fig:Figure_Isotropic_unloading_compressibility}). The variation of normal effective stress over time during unloading is shown in Supplementary Fig.~S1(b). The sample undergoes isotropic unloading by gradually reducing the cell pressure after initial loading. Isotropic unloading compressibility is determined as the slope between volumetric strain \(\varepsilon_v\) and mean effective stress \(p'\). The simulation is conducted as a drained test, where pore volume changes are monitored. At the end of each test, the volumetric strain \(\varepsilon_v\) and corresponding differential stress \(q\) are recorded at various time points. The resulting values are \(E = 1.36 \times 10^{-6} \, \mathrm{Pa}^{-1}\) and \(\beta = 2.13\). The detailed results are shown in Supplementary Fig.~S1(c).

\subsection{Pore pressure evolution during breaching}
The evolution of excess pore pressure during breaching reveals distinct responses in dense and loose subaqueous granular systems, governed by the initial packing density and fluid--particle interactions. Fig.~\ref{fig:Figure_excess_pore_pressure2} presents the spatial and temporal variations across monitored points (A, B,  D, E, and F for the dense configuration shown in Fig.~\ref{fig:Figure_excess_pore_pressure2}(a), and G, H, J, K, and L for the loose configuration shown in Fig.~\ref{fig:Figure_excess_pore_pressure2}(d)), which were strategically selected to capture localized pressure dynamics.

In the dense configuration, wall removal initiates an abrupt drop in pore pressure, followed by a gradual rise as fluid redistributes and particles rearrange (Fig.~\ref{fig:Figure_excess_pore_pressure2}(b)). This transient suction phase results from the system’s resistance to compaction. Over time, pore pressure stabilizes, reflecting the balance between fluid expulsion and interparticle contact forces. At point D, the temporal evolution of excess pore pressure (Fig.~\ref{fig:Figure_excess_pore_pressure2}(c)), plotted in Lagrangian coordinates across multiple time intervals (0.01~s, 0.1~s, 0.4~s, and 0.8~s), shows an initial negative pressure that progressively increases toward a steady state. This behavior aligns well with analytical predictions, with minor deviations during the transient phase attributed to localized heterogeneities and nonlinear fluid--particle coupling.

In contrast, the loose configuration exhibits a sharp rise in pore pressure following wall removal, driven by void collapse and rapid fluid compression (Fig.~\ref{fig:Figure_excess_pore_pressure2}(e)). The high initial pressure reflects the system’s susceptibility to compaction, followed by redistribution and a gradual decline toward equilibrium. At point J, the temporal pressure profile (Fig.~\ref{fig:Figure_excess_pore_pressure2}(f)) begins elevated and stabilizes over time. While the steady-state matches analytical solutions well, significant deviations are observed during the transient stage, underscoring the complex dynamics of loosely packed systems. 

These comparisons confirm that the analytical solution, despite its simplifying assumptions, effectively captures the steady-state behavior of excess pore pressure for both dense and loose subaqueous breaching scenarios. The good agreement observed at later times across different monitored points supports the applicability of the poromechanical framework in predicting long-term pore pressure evolution. The transient mismatches highlight the complexity of fluid–particle interactions during early deformation, and the possible role of transient frictional behavior of the granular materials in response to perturbations \cite{kohler2023rate, grelle2010shear, lemos2004shear}. 

The observation that transient pore pressure does not align with the analytical soil-mechanics model during the initial flow phases can inform further development of submerged granular-flow models, such as the compressible \(\mu(J)\), \(\phi(J)\) formulation proposed by \citet{rauter2021compressible}. Rauter's model provides a compressible two-phase continuum description of submerged granular motion by coupling the Navier–Stokes equations with a \(\mu(J)\), \(\phi(J)\) rheology. The constitutive relations \(\mu(J)\) and \(\phi(J)\) incorporate dilatancy, pore-pressure coupling, and critical-state behavior, offering a continuum representation of dense, submerged granular flow. Incorporating quantitative relationships derived from our rheological fitting  and the transient pore-pressure responses observed in our CFD–DEM simulations could extend Rauter's model to capture the poromechanical feedbacks that control flow initiation, and transient particle–fluid interactions.

\section{Conclusion}

This study investigated the collapse and runout of granular materials in subaerial (dry) and subaqueous (fluid-saturated) settings, focusing on the roles of initial packing density and fluid–particle coupling. Three-dimensional CFD–DEM simulations showed that dense packings formed narrower shear zones and shorter runouts due to greater resistance to deformation, whereas loose packings deformed more broadly and ran out farther because compaction is easier.

In subaqueous conditions, these trends were amplified: dense materials failed gradually via dilation and the generation of negative pore pressures, while loose materials failed rapidly due to compaction and transient positive pore pressures. Porosity evolution controlled deformation, fluid migration, and mechanical stability. Dense systems dilated during shear, drawing in fluid and increasing resistance; loose systems compacted, expelling fluid and reducing stability. These behaviors were governed by the strain rate and local stress state. Rheologically, the subaerial and subaqueous models under dense and loose conditions revealed how friction, strain rate, and the inertial (\(I_n\)) and viscous (\(I_v\)) numbers interact. Both environments exhibited overall rate strengthening, but for different reasons: in subaerial flows, friction increased modestly with \(I_n\) due to enhanced contact networks and force-chain persistence, whereas in subaqueous flows, frictional evolution reflected competing effects of fluid drag, pore-pressure buildup, and lubrication. Dense materials tended to maintain nearly constant porosity at low \(I_n\) and \(I_v\), indicating a dynamic balance between compaction and dilation, while loose materials showed larger porosity changes and greater frictional variability at higher strain rates. These trends highlight the sensitivity of granular strength to both fluid presence and initial packing fraction.

Simulation results agree with analytical predictions of excess pore pressure as flow approaches steady state, demonstrating that the model captures the coupling between fluid flow and granular deformation. Early-time discrepancies, however, underscore the complexity of fluid–particle interactions during transients. Overall, granular failure dynamics are highly sensitive to initial structure and hydromechanical coupling. These findings inform efforts to mitigate natural hazards such as landslides, sediment transport, and seabed instabilities. To enhance predictive capability, future work should incorporate time-dependent effects, spatial heterogeneity in material properties, and systematic comparison with field-scale observations. Moreover, because transient frictional behavior in granular materials may contribute to the early-time mismatch between analytical predictions and the CFD–DEM evolution of pore pressure, future models should explicitly include transient (rate-and-state) friction and be evaluated against analytical solutions that incorporate this mechanism.

\section{Supplementary Material}

See the supplementary material for additional figures and explanatory content detailing the poromechanical behavior of the modeled granular systems. Included materials cover the temporal evolution of permeability, derivation of isotropic unloading compressibility and dilation strength parameters, and boundary conditions applied for analytical model comparison. These supplemental results offer further insight into the influence of confining pressure, pore structure, and unloading on fluid migration and material response during breaching.

\section*{Author Declarations}
The authors declare no conflicts of interest.

\section*{Acknowledgments}
The research has been supported by start-up funds from the Department of Civil and Environmental Engineering and the Cullen College of Engineering at the University of Houston, as well as the American Chemical Society (ACS) Petroleum Research Fund under ACS-PRF Grant 66470-DNI8. We also thank Prof. Ruth Cardinaels, Associate Editor of Physics of Fluids, and the anonymous reviewers for their constructive feedback, which improved the clarity and rigor of this manuscript.

\section*{Data Availability}
The data that support the findings of this study are openly available in Zenodo at https://doi.org/10.5281/zenodo.16415581, Ref.~\onlinecite{ferdowsi_2025_16415581}.


%
%

%


\bibliographystyle{aipnum4-1}
\bibliography{POF_manuscript}

\end{document}